# Diffusing molecules share the 1/3 shot noise suppression of quantum conductors.

Henri Vo Van Qui[1,*], Ignacio Madrid[1,2], Simon Grall[3], Thibaut Jonckheere[4], Akira Fujiwara[5], Laurent Jalabert[1], Masayuki Hashisaka[6], Christophe Demaille[7], Soo Hyeon Kim[2], Nicolas Clément[1,*]

[1] LIMMS-CNRS/IIS IRL 2820, The University of Tokyo; Tokyo, Japan. [2] Institute of Industrial Science, The University of Tokyo; Tokyo, Japan. [3] LAAS-CNRS UPR 8001; Toulouse, France. [4] Aix Marseille Univ, Université de Toulon, CNRS, CPT, Marseille, France. [5] NTT Basic Research Laboratories; Atsugi, Japan. [6] Institute for Solid State Physics, The University of Tokyo; Kashiwa, Japan. [7] ITODYS-CNRS UMR 7086; Paris, France.

*Corresponding authors: hvvq@iis.u-tokyo.ac.jp, nclement@iis.u-tokyo.ac.jp

**Electrical noise measurement, especially shot noise, which arises from the discreteness of charge, is an indispensable tool in quantum transport for probing the nature of charge carriers beyond simple conductance measurements[1-4]. Diffusive conductors share a universal shot noise reduction of 1/3, a result obtained independently from quantum scattering theory[5], semi-classical kinetics[6,7], or by classical exclusion processes[8]. Until now, however, experimental tests of this universality have been limited to cryogenic conditions[9-12], leaving its origin incompletely understood. Here we show that ferrocene redox cycling in a microfluidic gap is a room temperature realization of this universality. We derive the full counting statistics of diffusing single-electron molecular shuttles and verify the predicted current noise experimentally, showing that the universal 1/3 shot noise suppression is recovered in the diffusion-limited regime. Our result identifies diffusion and sequential charge transfer as sufficient ingredients for this universal noise reduction, rather than quantum coherence, fermionic statistics or cryogenic conditions. We anticipate that this study will establish electrochemical microfluidics as a room-temperature platform for mesoscopic counting statistics and bring noise-based probes to molecular transport and reaction kinetics. Furthermore, this liquid-based quantum-inspired study will provide a novel insight on the yet to be understood links between quantum and biology[13].**

Because of the discrete nature of particles involved in charged transport, current flowing through a conducting channel fluctuates and gives rise to shot noise. This noise is quantified by the Fano factor $F = S^I/2eI$, the ratio of the noise spectral density to the Poissonian value corresponding to uncorrelated transfer of charge $e$. When the charge transfer events are independent, the process is Poissonian and $F = 1$, but when those events are correlated, $F < 1$ and the noise is reduced. Given that shot noise is also sensitive to the charge of the carriers, noise measurement has become an indispensable tool in quantum transport[1,14]. It gave the direct observation of the fractional charge of Laughlin quasiparticles in the fractional quantum hall effect[2,15] and of the double charge of Cooper pairs at a normal-metal/superconductor contact[16]. It was also used to detect the pairing in the cuprate pseudogap above the superconducting transition[3] and to probe strange metal behaviour, providing evidence that charge transport is not carried by conventional quasiparticles[4].

One of the most interesting and still debated topic in this field is the shot noise suppression in disordered conductors. Noise can provide information on a system beyond standard conductance measurement and in conductors where electron motion is diffusive, the full shot noise value is predicted to be reduced by a factor $F = 1/3$. This result has been obtained by several theoretical approaches applied to different mesoscopic models of diffusive conductors, revealing a surprising universality (Fig. 1). The reduction of the shot noise was demonstrated by Beenakker and Büttiker using a scattering matrix approach based on the Landauer-Büttiker formalism[5] (Fig. 1A). Nagaev obtained the same factor using a semi-classical Boltzmann-Langevin approach[6,7] (Fig. 1B), proving phase coherence is not required for the reduction of the shot noise. Derrida recovered this result for classical particles using a symmetric simple exclusion process (SSEP) along a 1D chain[8] (Fig. 1C).

Interestingly, not only the noise exhibits a universal reduction but also the higher order cumulants[8,17]. The suppression factors are the same across the different theories, hinting towards a unifying concept, whose origin lies beyond classical or quantum physics. Experimentally this reduced shot noise has been observed for several extended mesoscopic conductors with diffusive electrons transport[9-12], all at temperatures below 10K. As the temperature is increased, inelastic electron-phonon scattering dominates and are described by the Johnson-Nyquist noise. Therefore, a different platform is required at room temperature. Theories of first-passage time for random walks under confinement, with absorbing and reflecting boundaries, have considered how those boundaries and the starting point change the first-passage time statistics relative to free diffusion[18,19], yet that community and mesoscopic physics barely interacted. The reasons being the different targeted applications: diffusion-limited reactions[20] and search processes in biology versus quantum devices; and the room-temperature ceiling: inelastic scattering was assumed to restore the Johnson-Nyquist noise. Electrochemistry combines a Brownian single-electron cargo with interfacial electron transfer, as in redox cycling[21] (Fig.1D), but it was not considered as a mesoscopic physics problem either. Numerical simulations of redox cycling found a shot noise reduced by about a factor 1/3, without an analytical model[22], and experiments have typically remained about ten times above the Poisson-shot noise limit[23,24]. One can therefore ask whether mesoscopic shot-noise theory can describe an electrochemical system despite its inherently inelastic motion? If so, does that system follow the universal 1/3 suppression of diffusive shot-noise, and can this be observed experimentally?

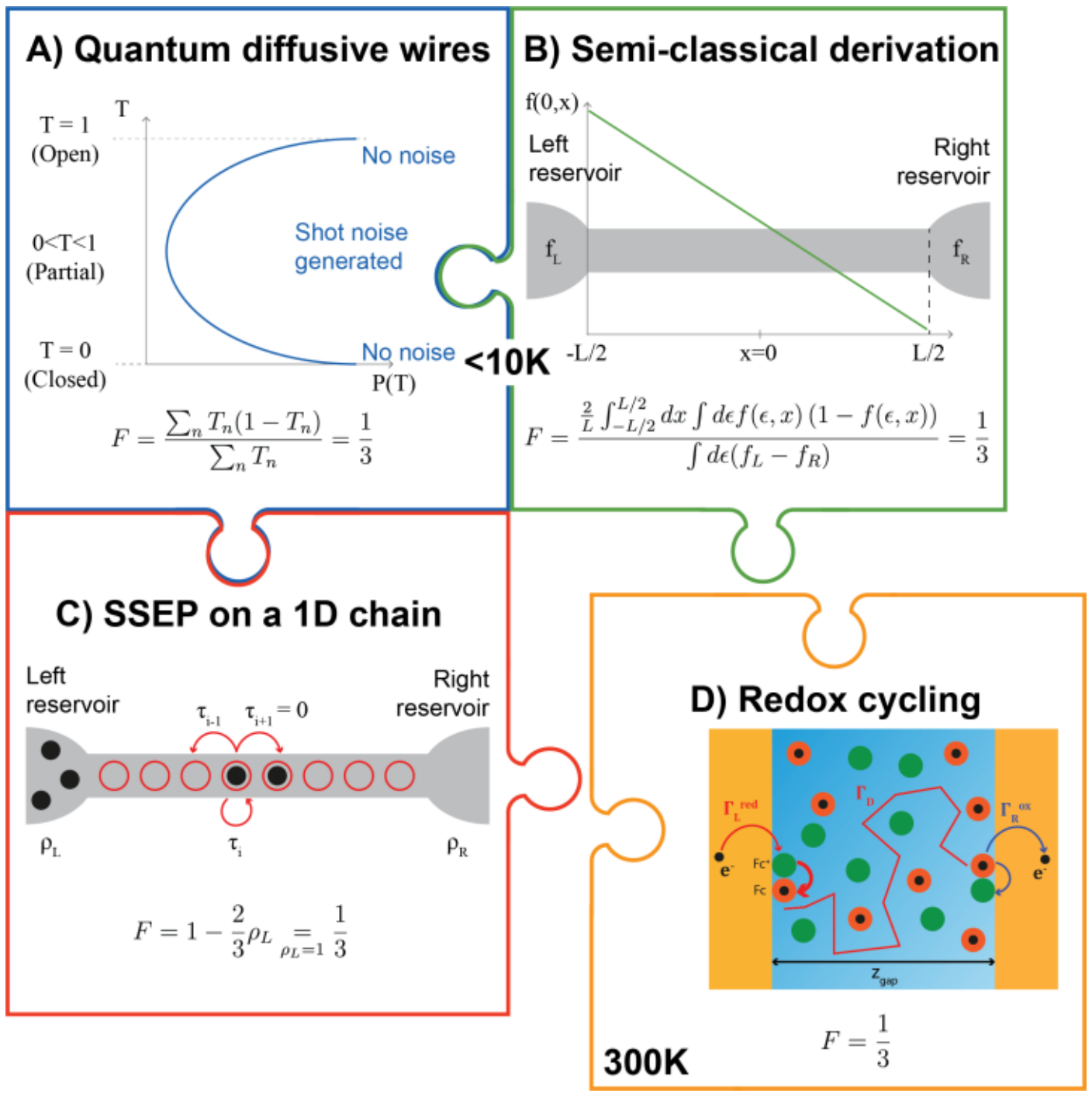

**Fig. 1. The three theories for universal 1/3 shot noise suppression and the proposed missing element.** (**A**) Quantum scattering in a diffusive wire in cryogenic conditions, $T$ is the transmission probability of an electron for a given conduction channel and $P(T)$ its distribution. (**B**) Semi-classical Boltzmann-Langevin kinetics. $f(\varepsilon, x)$ is the occupation at energy $\varepsilon$ and position $x$. $f_L$ and $f_R$ are the reservoir Fermi functions. (**C**) Symmetric simple exclusion process along a 1D chain. $\rho_L$ and $\rho_R$ are the probability densities of the left and right reservoir. (**D**) Redox cycling in a microgap at room temperature.

Here we provide a novel insight on this open question by presenting an analytical model and experimental results for noise in a closed electrochemical system. Electrochemistry offers an alternative setting that has barely been used to address this question because it was thought to be limited to cryogenic conditions. Redox cycling recovers a similar framework as the aforementioned theories, without the limitations inherent to a degenerate electron gas. The proposed model is bias-dependant and covers both micro- and nano-electrochemistry regimes. We confirm experimentally this bias-dependency and show it approaches the universal shot noise reduction of $1/3$ at high bias. Finally, we discuss this link between electrochemistry and mesoscopic transport as it holds for diffusing electrons and molecules, in cryogenic conditions as well as at room temperature.

**An analytical model of redox cycling noise.**

The first theoretical framework is based on Full Counting Statistics (FCS). This tool is commonly used in mesoscopic physics charge transport to extract the different cumulants describing the statistics of the charge carriers[25,26]. Here we apply it to an electrochemical system, where ferrocene (Fc) molecules act as room temperature single electron shuttles (Fig. 2A). We use the ferrocene/ferrocenium ($Fc/Fc^+$) as an archetypal one-electron, outer-sphere redox couple. To apply FCS to the redox reaction and diffusion of Fc in a microgap, we create a discrete $N$ site model to describe the Brownian trajectory of a single redox molecule together with stochastic electron transfers at the two biased electrodes (Fig. 2B). Because the electrolyte high ionic strength screens intermolecular electrostatics and the electric field in the gap, the molecules are effectively independent and we can consider more than $10^{12}$ of them in parallel at typical experimental concentrations ($c_B = 1$ mM Fc).

From this picture, we can construct the rate matrix $W(\chi)$ (Fig. 2C) of the counting statistics formalism[27] and obtain an analytical model. The model presented in Figure 2B uses $N = 3$ sites per charge state for illustration but the derivation of the analytical model is made in the continuous limit where $N \to \infty$ (Supplementary text). Unlike in the SSEP (Fig. 1C), here the electron cargo executes a round trip: it is oxidized at one electrode and reduced at the other. There is no exclusion in the bulk: sites are not blocked by occupancy but the charge-state exclusion acts only at the interface[21]. The hopping particles are not electrons on a fixed lattice but the molecule itself is the cargo that shuttles charge between the two electrodes, so the number of cargos is the number of $Fc/Fc^+$ molecules in the microgap. The diffusive hop rate is $\Gamma_D$ and is of the Fick form. The interfacial electron transfer rates are $\Gamma_{ox}^L$, $\Gamma_{red}^L$, $\Gamma_{ox}^R$ and $\Gamma_{red}^R$ are of Butler-Volmer form[28-30], with a transfer coefficient $\alpha = 0.5$. This is the analogue of a Landauer formalism whose broadening is set by the Franck-Condon reorganization energy of the polar solvent (here water), rather than by a coherent level width. As for a tunnel current, those rates decay exponentially with the distance of the molecule from the electrode. Electrons are counted at the left electrode (anode), on the bond marked $e^{\pm i\chi}$, where $\chi$ is the counting field of full counting statistics[27].

The mean current and the zero-frequency noise are the first two derivatives of the dominant eigenvalue $\mu(\chi)$ of $W(\chi)$ at $\chi = 0$, the Fano factor is their ratio[27]. Once the Butler-Volmer and Fick rates are given as functions of the anode bias $V$ (versus Ag/AgCl), $I(V)$ and $S^I(V)$ follow (Fig. 2D, E). Experimentally, the cathode is usually held at a potential much lower than the formal potential of the Fc/Fc$^+$ couple: $E_0 = 0.25$ V versus Ag/AgCl, in that case we obtain:

$$I(V) = I_{lim}\frac{1}{1 + e^{-f(V-E_0)} + \Lambda e^{-f(V-E_0)/2}} \qquad (1)$$

$$S^I(V) = 2eI(V)F(V) \qquad (2)$$

where $I_{lim} = FADc_B/z_{gap}$ is the diffusion limited current plateau and $\Lambda = D/k_0 z_{gap}$ is the dimensionless rate constant with $D$ the diffusion coefficient, $k_0$ the standard heterogeneous electron transfer rate constant and $f = F/R_gT = e/k_BT$. For the Fano factor:

$$F = 1 - \frac{2}{3}\rho_{ox}^L - \frac{\Lambda\rho_{ox}^L}{3}\left(\sqrt{\Lambda^2{\rho_{ox}^L}^2 + 4\rho_{ox}^L(1-\rho_{ox}^L)} - \Lambda\rho_{ox}^L\right) \qquad (3)$$

where $\rho_{ox}^L = I/I_{lim}$ is the oxidized fraction at the anode. $\Lambda \ll 1$, is the condition for the Nernstian regime where electron transfer is fast compared to diffusion. In that case, $\rho_{ox}^L$ takes the Fermi-Dirac form and Equation 3 simplifies to (Fig. 2F):

$$F = 1 - \frac{2}{3}\rho_{ox}^L \qquad (4)$$

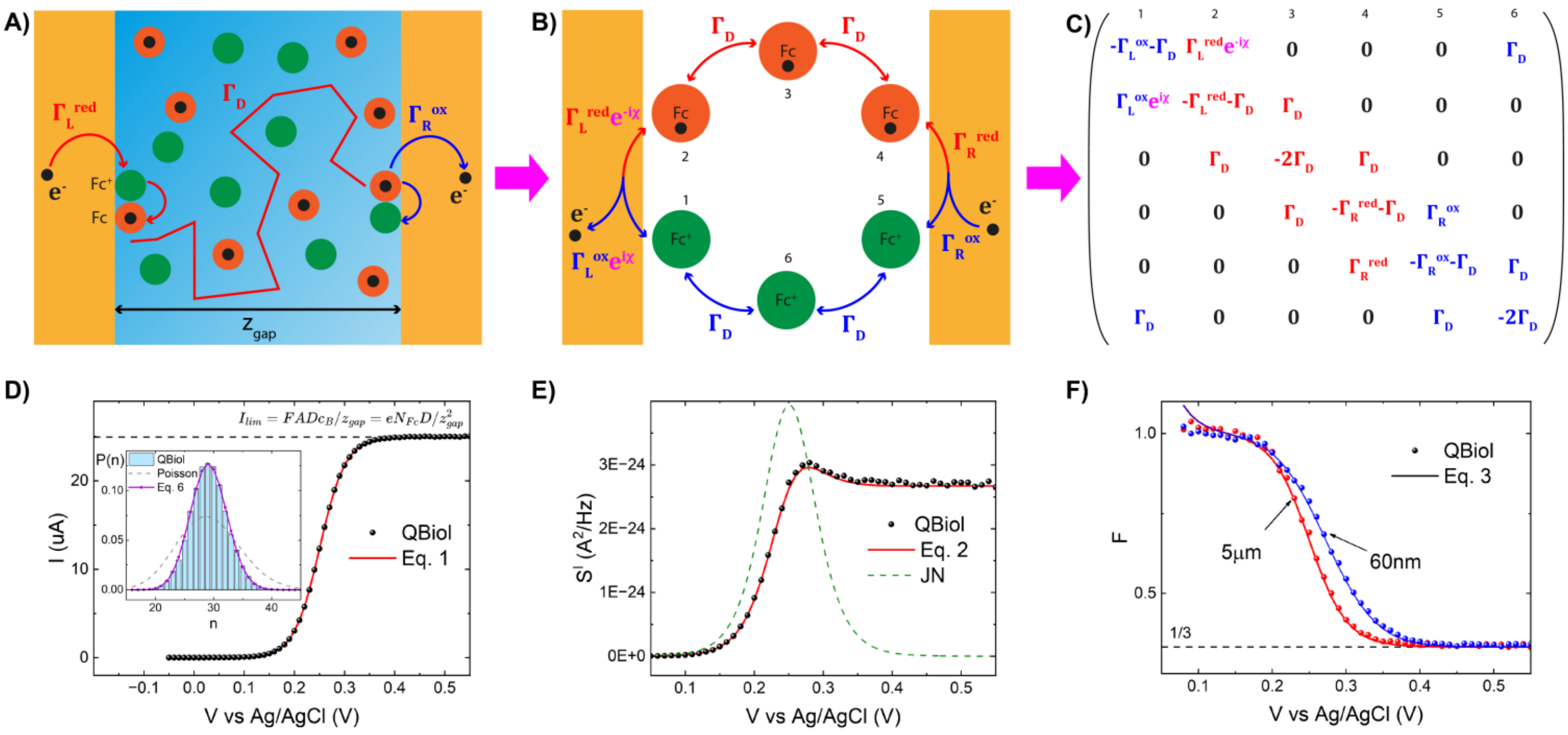

**Fig. 2. Full counting statistics electrochemical noise model.** (**A**) Schematic representing the Fc molecules diffusing inside a microgap between two electrodes. (**B**) Discretized model ($N = 3$) illustrating the electronic transfer and diffusion rates. (**C**) Equivalent rates matrix derived from the discretized model. (**D**) Theoretical current calculated from the rates matrix compared with the digital twin experiment by QBiol for a gap $z_{gap} = 5$ µm, an electrode area $A = 2$ mm$^2$, a concentration $c_B = 1$ mM and by taking $D = 6.5\text{e-}10$ m$^2$/s for the diffusion coefficient and $k_0 = 0.004$ m/s for the heterogeneous rate constant. Inset shows the full counting distribution from QBiol, fitted with Equation 6 (purple curve). (**E**) Theoretical noise compared with QBiol and the Johnson-Nyquist (JN) contribution $4k_BT\delta I/\delta V$. (**F**) Fano factor reaching the universal $1/3$ limit in the diffusion limit, for a Nernstian regime ($z_{gap} = 5$ µm) and non-Nernstian regime ($z_{gap} = 60$ nm).

This corresponds to the Fano factor of the SSEP[11] (Fig. 1C) for an emp ty right reservoir. Thus, the redox-cycling FCS and the SSEP agree not only at $F = 1/3$ ($\rho_{ox}^L = 1$), but for every $\rho_{ox}^L$ along the voltametric wave in the Nernstian regime. Analytical $I(V)$, $S^I(V)$ and $F(V)$ from Equations 1 to 3 are confirmed by QBiol (Fig. 2D-F), a stochastic simulator of molecular electron transfer[31,32]. For this study, QBiol was extended to full counting statistics: each realization records the integer transfer count, from which the current and the noise follow (Fig. 2D, inset). The agreement between the analytical model and QBiol across the voltametric wave shows that this account of sub-Poissonian redox-cycling noise, including the universal $F = 1/3$, holds even though the gap operates in the inelastic regime. The mean current also matches what is established for redox cycling in nanogaps, including in the non-Nernstian regime at $z_{gap}$ of tens to hundreds of nanometers[33]. The present model adds the noise. This non-Nernstian regime is a concrete case in which $F(V)$ departs from Equation 4 while still converging to $F = 1/3$ at high bias. $F$ falls from the Poisson value toward $1/3$ within a few $k_BT$ of the formal potential $E_0$. This is different from a metal, where the crossover out of the thermal regime[14] is set by $eV \gg k_B T$. $S^I$ is not the Johnson-Nyquist $4k_BTG$, with $G=\delta I/\delta V$, except at zero net current (Fig. 2E, Fig. S6). Along the voltametric wave, the two contributions diverge: $S^I$ lies below $4k_BTG$ on the rising part, but remains finite on the plateau where $G \to 0$.

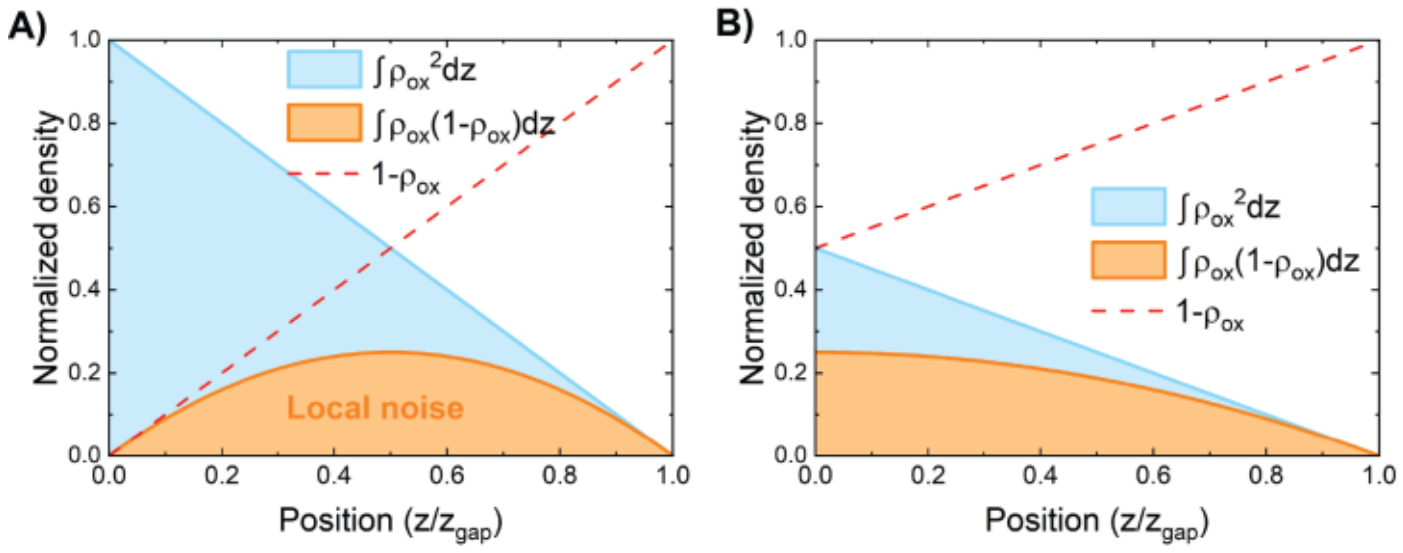

**Fig. 3. Boltzmann-Langevin analytical model for the electrochemical noise.** Analytical oxidized (blue line) and reduced (dashed red line) profiles along the gap for redox cycling in the Nernstian case. The orange line is the occupancy/noise kernel $\rho(z)\big(1-\rho(z)\big)$. In this formalism, F is the ratio of the orange to the blue areas. (**A**) $\rho_{ox}^{L}=1, \rho_{ox}^{R}=0$ (**B**) $\rho_{ox}^{L}=\rho_{ox}^{L}=0.5$.

Although the redox-cycling FCS recovers Equation 4 and the $F=1/3$ of Figure 1, the Fano factor alone is not enough to claim the universality. For instance, graphene at the Dirac point also yields the same Fano factor, but from pseudodiffusive (evanescent-mode) transport at the Dirac point rather than from a disordered wire[34]. A similar number is approached at room temperature in atomic-scale gold contacts, where many Landauer channel mix[35]. The three approaches presented in Figure 1 share more than a common Fano factor as all the higher order cumulants of the transferred charge also match (including noise, skewness and kurtosis). We can now test whether the electrochemical gap obeys that law by deriving the cumulant generating function of this electron shuttle from a probabilistic first passage approach, complementary to the FCS derivation in the Nernstian regime (Supplementary text):

$$S(\chi)=\frac{\langle n\rangle_t}{\rho_{ox}^{L}}\left(\operatorname{arcsinh}\sqrt{\rho_{ox}^{L}(e^{i\chi}-1)}\right)^2 \qquad (5)$$

Where $\langle n\rangle_t$ is the mean count of electrons being transferred during a large enough time window $t$. In the Nernstian and $F=1/3$ limit it coincides with Levitov's full counting statistics of a diffusive quantum wire[27], with a cascade of Langevin sources[17], and with the open SSEP for every boundary occupancy[8]. The inset in Figure 2D shows a very good agreement between the full counting distribution from QBiol and the probability distribution $P(n)$ derived in the same manner as in single-electron counting on semiconductor quantum dots[36]:

$$P(n)=\frac{1}{2\pi}\int_{-n}^{n} d\chi e^{S(\chi)-in\chi} \qquad (6)$$

From this counting and cumulant generating function, we can also show that the Fano factor, the skewness and the kurtosis versus $\rho_{ox}^{L}$ fall from their Poisson value toward $1/3$, $1/15$ and $-1/105$ respectively, the universal values for diffusive systems (Fig. S8). In other words, the counting statistics of the transferred charge is the same in a diffusive quantum wire at cryogenic temperature and in an electrochemical microgap, even though in the latter the cargo is a ferrocene molecule cycling at room temperature.

To see how this universality appears in the local noise language of Figure 1B, we evaluate the Fano factor from Nagaev's Boltzmann-Langevin theory[6], without integrating over a degenerate electron gas: only a single redox occupancy enters. The blue and red curves in Figure 3 are the usual oxidized and reduced profiles along the gap; the orange curve is the local noise. Figures 3A and B locates the $F=1/3$ suppression in space. For $\rho_{ox}^{L}=1$ and $\rho_{ox}^{R}=0$, the local noise source $\rho_{ox}(1-\rho_{ox})$ vanishes at both electrodes and peaks at mid-gap. The Fano factor is the orange area divided by the blue area, equal to $1/3$. Figure 3B shows the same construction at $\rho_{ox}^{L}=0.5$: the noise source is strongest near the oxidizing electrode, and the area ratio is $2/3$. Both cases follow Equation 4. That spatial representation extends to more complex geometries, where the occupancy profile is obtained by finite-element calculation. The $F=1/3$ suppression itself, however, holds only in a closed gap. An ultramicroelectrode will give full shot noise, even though the $I(V)$ and $\rho_{ox}$ profiles can look similar[22].

All the models described above assume the zero-frequency limit. Indeed, the diffusion coefficient of a redox molecule is several orders of magnitude lower than the electronic diffusion coefficient in a metal, so the frequency dependence must be considered experimentally. For $f \gg 2D/z_{gap}^2$, the correlations are lost and full shot noise is expected[22,33].

## Measuring redox-cycling shot noise in a microgap

We explore here the room temperature reduction of the shot noise due to the diffusion of ferrocene dimethanol ($Fc(MeOH)_2$) molecules in a microgap. We use these molecules because they are more soluble in aqueous solutions, but this doesn't affect the underlying physics as described by the redox cycling analytical model. The Fc concentrations used here are 1 mM and 2 mM in an $NaClO_4$ solution at 0.5 M, which acts as the electrolyte and helps screen the electric field at the electrode. For this counter ions concentration, the Debye length is ~0.43 nm. These Fc concentrations were chosen to have sufficient levels of current while remaining below the solubility limit.

To overcome the challenges of room temperature noise measurement a custom probing stage setup (Fig. 4A) was built combining microfluidics and precision electronics. A particular attention was given to grounding and shielding as well as the optimization of the different instruments. The microfluidic device consists of patterned platinum electrodes on quartz wafers on either side of a gap of well-defined thickness of 5 µm. For this gap size, $\Lambda \approx 0.03$ corresponding to the Nernstian regime, validating the use of the simplified Equation 4 to describe the experimental Fano factor. The Fc molecules can freely move inside this enclosed gap and the current and current noise are measured. As expected, the current follows Equation 1, a Fermi-Dirac wave that saturates at a diffusion-limited plateau $I_{lim}$ at high bias (Fig. 4B). The level of the plateau doubles from 1 mM to 2 mM Fc concentration.

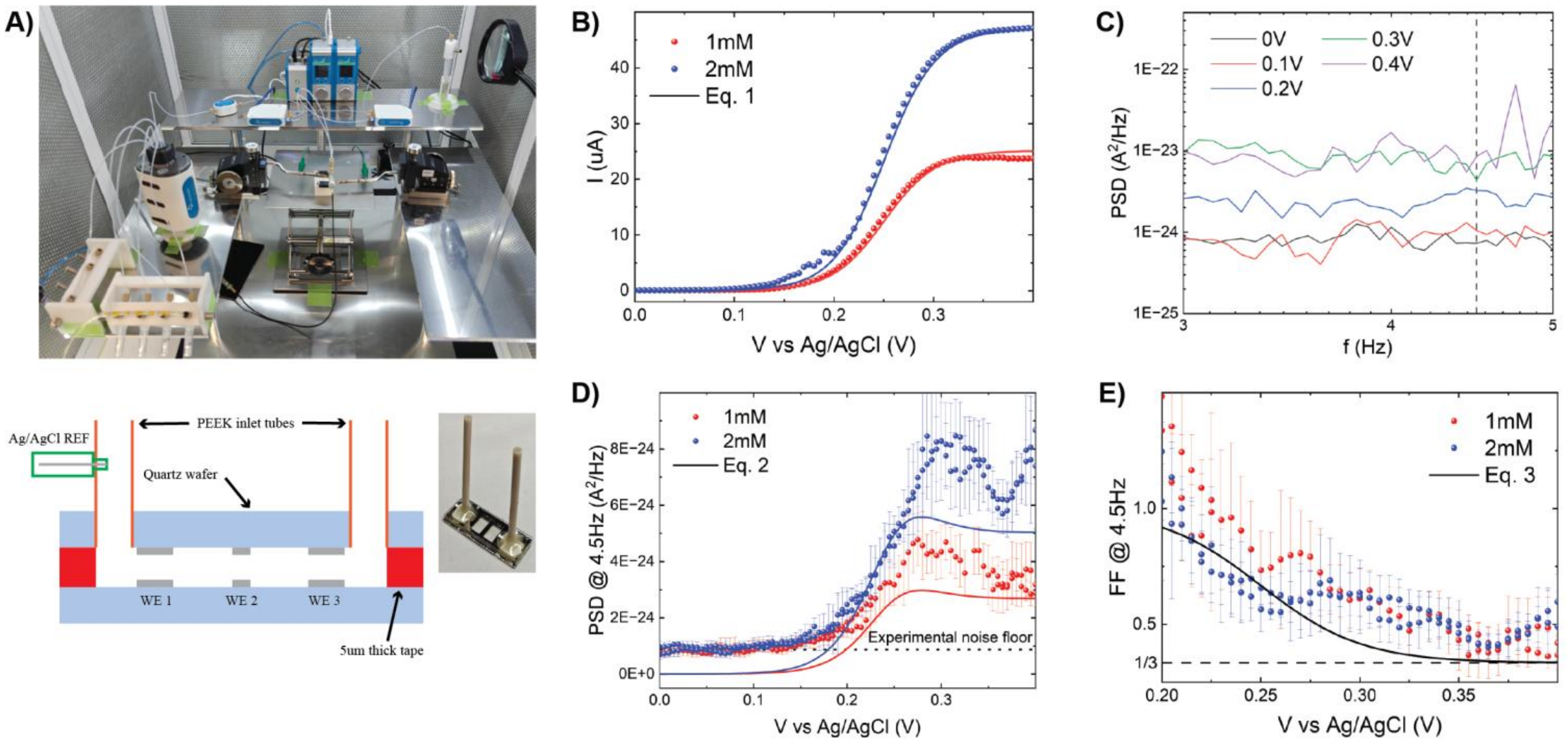

**Fig. 4. Room temperature redox cycling current and noise experiment.** (**A**) Picture of the experimental setup combining microfluidics and noise measurement, schematic and picture of the micro gaped device. (**B**) Experimental redox cycling current obtained with the same parameters as Fig. 2D-F at two different concentrations of $Fc(MeOH)_2$ and compared to the analytical model (Eq. 1) for 1 mM (solid red line) and 2 mM (solid blue line). (**C**) Noise spectrum for 2mM $Fc(MeOH)_2$. (**D**) Experimental power spectral density noise at 4.5Hz compared to the analytical model (Eq. 2) for 1 mM (solid red line) and 2 mM (solid blue line). (**E**) Experimental Fano factor at 4.5 Hz compared with the concentration-independent Equation 4. The $F = 1/3$ level is indicated.

The noise spectrum lies well above the instrumental floor, but only a limited frequency window is usable for shot noise study in this device (Fig. S4). At low frequency (below 1 Hz), a $1/f^2$ or $1/f^3$ background appears, previously linked to adsorption/desorption[24] of $Fc(MeOH)_2$ or $Fc^+(MeOH)_2$. At high frequency (above 10 Hz) the noise rises, which we attribute to dielectric polarization noise of the double layer capacitance. In between, the spectrum is white and bias-dependant (Fig. 4C). The experimental noise $S^I(V)$ is well described by the theoretical model (Fig. 4D): it is bias-dependant, presents a slight overshoot near the formal potential $E_0$ and displays a plateau in the diffusion-limited regime. We were able to get this experimental confirmation of the model for two different concentrations which is an extra degree of freedom compared to the metallic case. Indeed, the ability to modify the density of charge carriers is specific to electrochemistry and as expected, the Fano factor is independent on this parameter. It follows Equation 4 and, in the diffusion-limited regime at high bias, approaches $1/3$, the universal shot noise reduction value for diffusive conductors (Fig. 4E).

## Discussion and perspective

The general equation for the Fano factor (Eq. 3) depends on the parameter $\Lambda$ which is standard in micro- and nano-electrochemistry, but new for universal shot noise. When $\Lambda \ll 1$, it reduces to the simplified equation (Eq. 4), which coincides with Derrida's SSEP result[8]. By integrating noise in space following a Boltzmann-Langevin approach we can recover the same simplified equation but this approach cannot give the general equation without significantly complexifying the model.

Noise suppression originates from correlations whose exact origin, including Pauli exclusion or the bulk exclusion of the SSEP has been a long-standing question. In redox-cycling both are absent, yet $F = 1/3$ remains. We suggest that diffusion is the common ingredient, and that different kinds of exclusion and correlation can yield the same counting statistics. In this study, the key component is a molecule carrying a single charge (only one site occupied on the ox/red ring at any time) and the internal correlation in its round trip due to the sequential nature of the process: the return trip will not start until electron transfer occurs at the right electrode after the forward trip.

Beyond mesoscopic physics, our result raises a simple question: when does an observation prove that a system is quantum? This question is central in quantum biology, where some effects are quantum only in a trivial sense, because they set energy levels or reaction rates, while others would rely on quantum coherence[13,37]. Classical models have reproduced oscillations initially considered as evidence of coherent energy transfer in photosynthesis[38,39]. Redox cycling offers a simple example of this situation. Electron transfer at each electrode is a tunnelling event, quantum only in the nano electronic sense, but the molecule crosses the gap by classical Brownian motion. The water acts only as a thermal bath, unlike at nanoscale carbon interfaces, where water flows couple to the electronic excitations of the surface[40,41]. At high bias, the counting statistics are indistinguishable to those of a quantum diffusive wire at a few Kelvins. The $1/3$ suppression, and even the full counting statistics, therefore reflects diffusion rather than quantum mechanics. Redox cycling can thus serve as a classical baseline: in molecular or biological electron transfer, departures from it could reveal effects beyond simple diffusion, such as quantum coherence or transport by unconventional quasiparticles[4].

The Johnson Nyquist estimate differs sharply from the metal case: near the formal potential it can even exceed the measured current noise. The two coincide at zero current, where the fluctuation-dissipation theorem holds, as in the electrochemical noise of a redox monolayer[42]. Nevertheless, because the differential conductance $G \to 0$ at high bias, $4k_BTG$ vanishes and the shot noise in redox cycling can be read directly, without the need of subtracting the Johnson-Nyquist contribution as is typically necessary in metals. This remains possible even though the motion in the gap is inelastic. In a metal, electron-phonon scattering restores Johnson Nyquist noise.

Experimentally, a 5 μm gap is a useful compromise as it is large enough that the interfacial backgrounds (adsorption-desorption and dielectric polarization noise) do not require specialized electrode engineering, yet small enough that a white-noise window remain below the diffusion frequency $f_D = D/z_{gap}^2$. Typical electrochemical setups, including single-molecule devices, remain about ten times above the Poisson shot-noise[23] but this floor is not fundamental. Surface engineering, together with experimental techniques from quantum devices could cut the background by at least an order of magnitude. That would raise the signal-to-noise ratio of single-molecule electrochemical devices. The same isolation of the shot noise would also give an independent test of open questions in biomolecular systems, including ballistic Brownian motion[32,43,44], reorganization energy and non-ergodic sampling of molecular configurations[45]. With a different probe or solvent, the same noise analysis could be extended to the non-Markovian regime[46].

This room-temperature demonstration of shot-noise suppression bridges mesoscopic physics and electrochemical transport, and establishes noise analysis as a new probe of chemical reactions and molecular dynamics, opening a new avenue in electrochemistry.

**Acknowledgements:**
The device fabrication was conducted in Takeda Super Cleanroom, Centre of The University of Tokyo for The Advanced Research Infrastructure for Materials and Data Hub, with the help of S. Li and H. Dai. We thank T. Misawa for the quartz wafer processing; D. Bourrier and A. Beghersa from LAAS-CNRS for the fabrication of the Cr mask used for the Pt deposition; Y. Li for the AFM picture and C. Gerbelot, T. Martin and J. Rech for the useful discussion about the models. **Funding statements:** This work was supported by the French "Agence Nationale de la Recherche" (ANR) through the HYPOSEL and ESHOT projects, by the "Mission for Interdisciplinary and Transverse Initiatives" (MITI) of the CNRS through the BIOSTAT project and by the "Advanced Research Infrastructure for Materials and Nanotechnology in Japan" (ARIM) of the Ministry of Education, Culture, Sports, Science and Technology (MEXT), Grant Number JPMXP1226UT1074. **Authors contributions:** H. V.V.Q. developed the theory, fabricated the devices, designed the acquisition system, conducted the experiments and analysed the data; I. M. developed the theory; S. G. conducted the simulations; T. J. developed the theory; A. F. and M. H. contributed to the scientific interaction on noise theory and measurements; L. J. designed the acquisition system; C. D. contributed to the scientific interaction on electrochemistry; S.H. K. contributed to the scientific interaction on microfluidics and supervised the project; N. C. conceived and supervised the project. The paper was written by H. V.V.Q. and N. C. All authors actively participated to the discussions on the paper. **Competing interests:** The authors declare that they have no competing interests. **Data, code and materials availability:** All data are available in the manuscript or the supplementary materials.

# Supplementary information

## 1. Computational and analytical methods

### 1.1 QBiol Brownian redox-cycling simulations

QBiol[31] was run in free-particle electrochemistry mode for a closed two-electrode gap. The QBiol data sets of the paper come from two campaigns: $z_{gap} = 5$ µm and 60 nm. The simulation temperature was 293 K and the diffusion coefficient was $D = 6.5\text{e-}10$ m$^2$ s$^{-1}$. The potential of the anode (left electrode) $V_L$ was swept from $-0.30$ to $+0.30$ V in 10 mV increments, the potential of the cathode (right electrode) was $V_R = -0.30$ V. QBiol considers the formal potential at 0 V to remain general. Then, for this study where Fc/Fc$^+$ couple is considered, we shift all the potentials by 0.25 V versus Ag/AgCl 3M KCl reference electrode. This value is typical for Fc/Fc$^+$ and corresponds to the experimental value in this study.

Brownian motion and interfacial transfer were generated on the fly (Butler–Volmer, $\alpha = 0.5$, $k_0 = 4.0\text{e}7$ s$^{-1}$, tunnelling decay $\beta = 10$ nm$^{-1}$). Each GPU thread is an independent single-molecule trajectory. Stored Brownian motion TRACK files written in the same runs were kept for checking diffusion properties. At $z_{gap} = 5$ µm, ten independent realizations were generated. Each used 1,091,840 independent GPU copies (4265 blocks × 256 threads) and a counting time window of 1.125 s at each bias. Error bars are the variability among these ten regenerated ensembles. At $z_{gap} = 60$ nm, a single simulation was used with 32,768 copies (128 blocks × 256 threads).

The reason why the smaller gap uses less threads is because the code is optimized for two different configurations: "large gaps" ($z_{gap} \geq 1$ µm) and "small gaps" ($z_{gap} < 1$ µm). Also, in the $z_{gap} = 60$ nm configuration, the counting time window is not homogeneous. It was 1.46 ms up to $V_L = -0.05$ V, and reduced down to 0.114 ms at 0.3 V, corresponding to $\langle n \rangle \sim 20$. In terms of simulation time, the 10 runs for the $z_{gap} = 5$ µm campaign was performed in 71.3 h (~3 days) and 2.73 h for $z_{gap} = 60$ nm (+ <10% of these times for Brownian motion TRACKS generation).

### 1.2 Counting distributions and cumulants

For every voltage, QBiol records the integer net count *n* at the left electrode. At 5 µm, each run is converted into a probability $P(n)$; the error bars are the standard error across the ten runs. The second cumulant use the same ten runs. The 3$^{rd}$ and 4$^{th}$ cumulants use five paired runs to increase the number of events and remove the few rare events that can affect the estimated value. At 60 nm, the histograms come from the single run, with no pairing.

### 1.3 Open access web version of QBiol

QBiol is an open access on-line Quantum Bioelectrochemical simulator (qbiol.org). The "Noise" module will be added to the actual online version of QBiol after publication of this article.

## 2. Experimental methods

### 2.1 Chemicals and solutions preparation

Ferrocene solution was prepared by dissolving ferrocene dimethanol $Fc(MeOH)_2$ crystals from Combi-Blocks at 97% purity in a 0.5 M $NaClO_4$ aqueous solution with DI water. Here $Na^+$ and $ClO_4^-$ ions act as the counter ions. $NaClO_4$ was purchased from Sigma Aldrich at 98% purity. Solutions are filtered and degassed for 15 minutes under $N_2$ prior utilization. The Fc solutions of 1 mM and 2 mM were prepared using a Sartorius high precision scale. $NaClO_4$ is expected to have a diffusion coefficient of 6.5e-10 $m^2/s$ [33].

### 2.2 Experimental setup

The device is made of two quartz wafers on which patterned 2nm Ti + 50nm Pt electrodes were sputtered. This allows for different configurations to be chosen for the measurement. Two holes were drilled in the upper wafer to connect the 1×15 $mm^2$ tape-defined channel to a Fluigent microfluidics setup used to inject the different liquids (Fc solution or cleaning solutions) inside the microgap. The thickness of the tape joining the two wafers is 5 µm. PEEK microfluidic tubes are glued to the holes with Masterbond glue optimized for PEEK materials, then cured at 65°C overnight. A microfluidic T-junction accommodates the Ag/AgCl reference electrode (ALS RE-3VT), and electrical contact is made with probe needles inside a Faraday cage. Liquid flow is stopped during every current-noise acquisition and the solution is trapped inside the gap using mechanical switches from Fluigent. As the pumps are pressure driven under $N_2$ gas, the inlet/outlet sealed, and top/bottom of the microfluidic device are made of quartz (not porous to gas as PDMS), the solution is stable in time and can be used for multiple days.

The electrical measurement setup is controlled by a LabVIEW program. The swept working electrode (anode) is driven by an NF LP6016 low-noise voltage source, in 5 mV increments. The redox current is measured at the cathode using a Stanford SR570 current pre-amplifier on a low gain setting. After the current reaches steady state at each voltage point, its DC component is cancelled with a second NF LP6016 combined with a Vishay metal foil 100 kΩ resistor to act as a low noise current source. Residual fluctuations are amplified with a gain of 1e7 V/A by the Stanford SR570 current pre-amplifier on Low Noise setting with an additional 6dB high pass filter with a cutoff frequency of 0.03 Hz to remove any residual DC component. In these conditions, the bandwidth is 2 kHz. The residual fluctuations are then digitized with a NI 4431 24-bit DAQ with a sampling rate of 1.024 kHz and 16,384 samples.

The time domain signal is converted to power spectral density by doing a Fast Fourier Transform using a Hanning windowing. Spectra are then averaged 20 times for each voltage point by RMS averaging. We then take the value of this PSD at a given frequency for each voltage point to get the voltage dependency. This is then averaged using a moving average technique over two neighbouring points. This also gives us the error bar on the PSD. A complete cyclic voltammogram is performed (a forward and reverse voltage sweep). Data corresponding to the reverse voltage sweep are used for this study. We notice that the overall experimental setup is more stable in the return sweep. The total experimental time per concentration was 25 h.

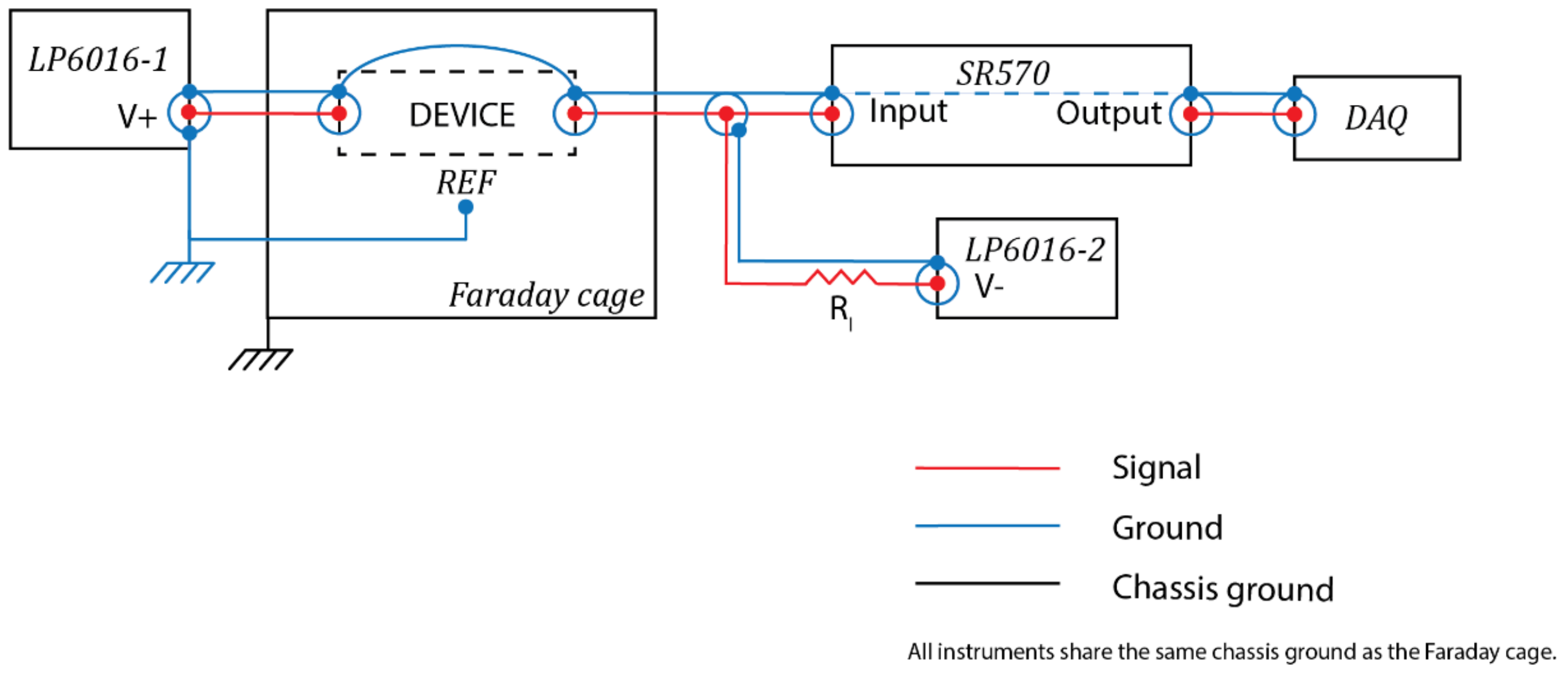

**Fig. S1. Schematic of the experimental setup showing the grounding and shielding configuration.**

### 2.3 Electrode roughness

The Pt surface RMS roughness of 1.48 nm (arithmetic roughness of 1.18 nm) is small compared with the 5 µm gap and is therefore negligible for the diffusion length, mean redox-cycling current and cargo shot noise.

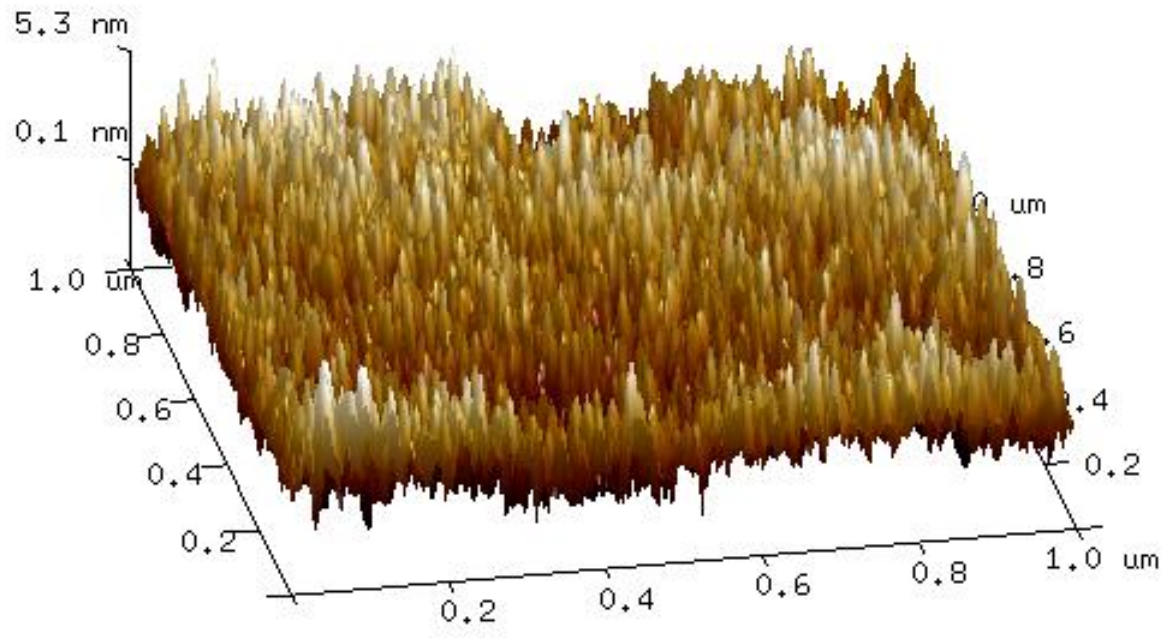

**Fig. S2. AFM height image of the Pt surface.** The AFM height image was corrected using an XY first order plane fit. Surface roughness was then calculated over a representative 1x1 $um^2$ region. The resulting RMS roughness was 1.48 nm and arithmetic roughness was 1.18 nm.

### 2.4 Instrumental background

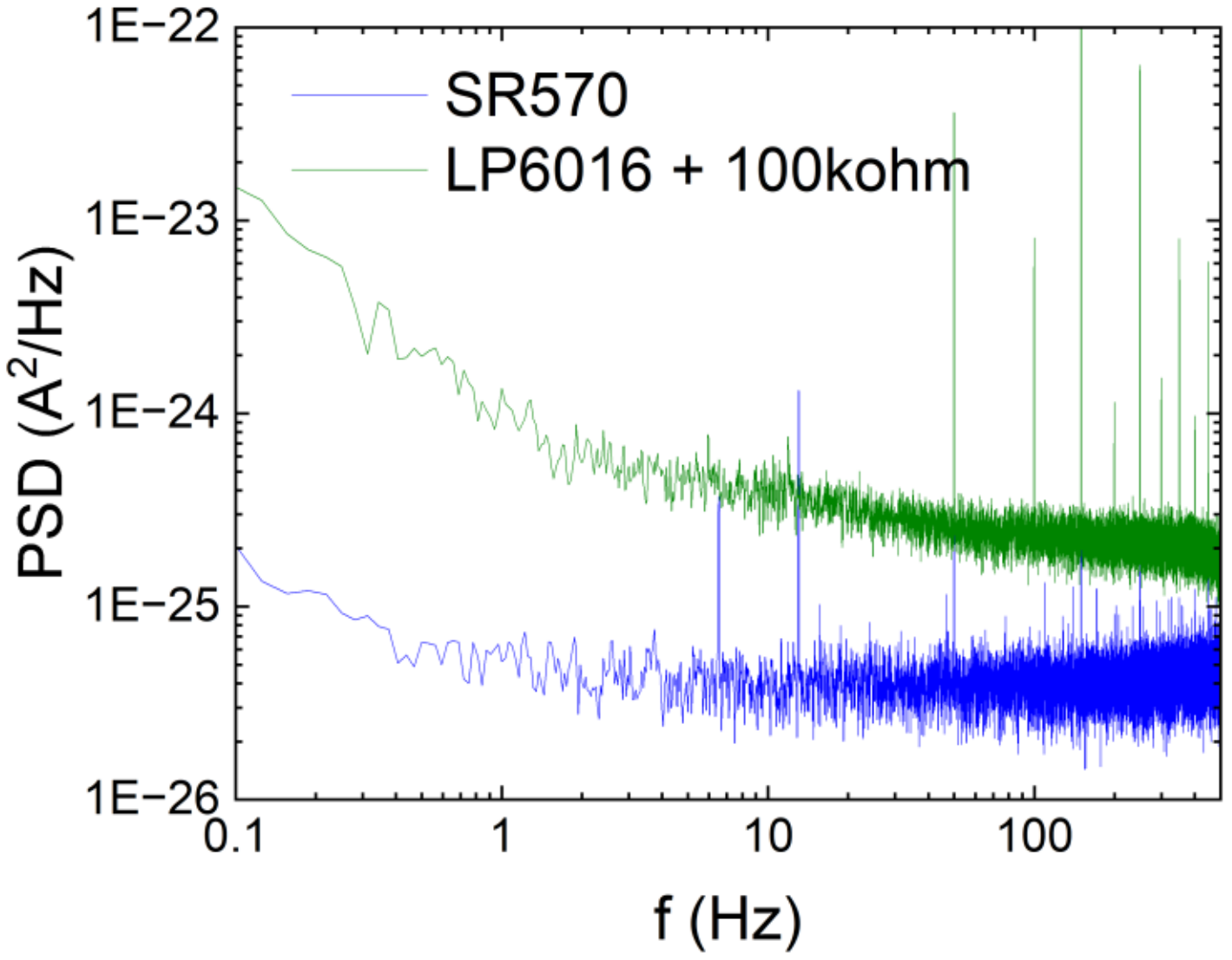

**Fig. S3. Background current noise for instruments.** The current pre-amplifier (Stanford SR570) was set with a gain of 1e7 V/A and Low Noise setting. The current source (NF LP6016 + 100kΩ resistor) was set to apply an offset up to -100 µA.

### 2.5 Noise spectrum summary

This summary shows that the measured noise is well above the instrumental noise. The noise at 0 V also corresponds to the noise measured on the same device with only $NaClO_4$ solution and corresponds to the experimental noise floor as indicated in Figure 4D. At high frequency, the noise seems to fit with the dielectric polarisation noise (red curve) which is of the Johnson-Nyquist form and originates from the interfacial double layer capacitance[47,48]. This noise is calculated from the measured impedance (Fig. S6). The noise for the NF LP6016 + 100kΩ resistor current source is the same as showed in Figure S3 and is in the case where it is used to apply an offset of -100 uA. When it is used to apply smaller current offsets, this noise is also smaller.

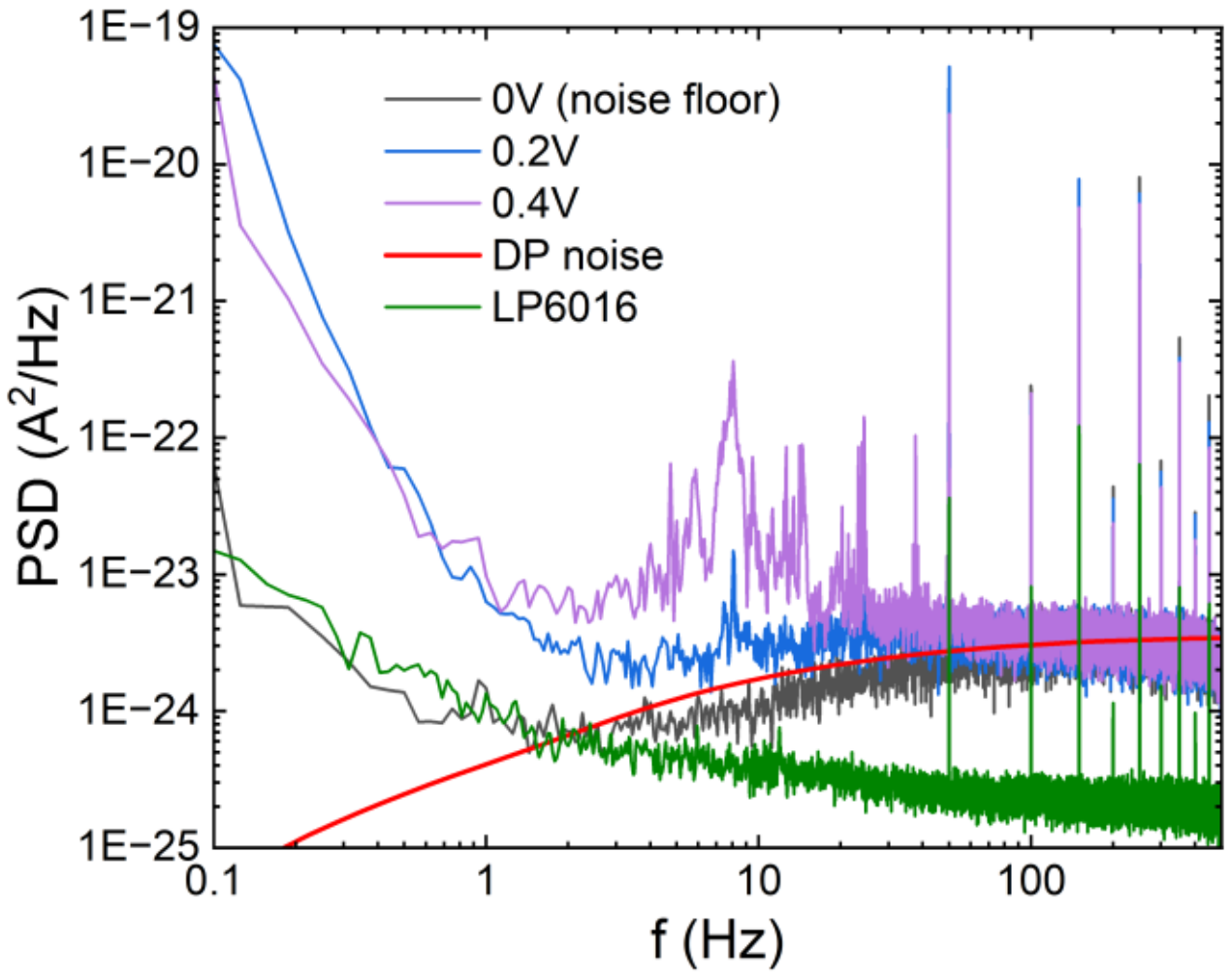

**Fig. S4. Noise spectrum summary.** PSD for Fc 2 mM for different biases from the same dataset as Figure 4C from the main text, noise spectrum from the NF LP6016 + 100kΩ resistor current source (green curve) which is the noisiest instrument and DP noise calculated from the measured impedance (red curve).

### 2.6 Impedance measurement

Impedance measurement was made to evaluate the real part of the impedance, which originates from the imaginary part of the double layer capacitance, and used to predict the Johnson-Nyquist contribution $4k_BT/\mathrm{Re}(Z)$ as described in Figure S4. To do so we used the Zurich Instrument MFLI on impedance measurement setting. The test signal was fixed to 0.3 V vs Ag/AgCl with the frequency swept from 0.1 to 2000 Hz.

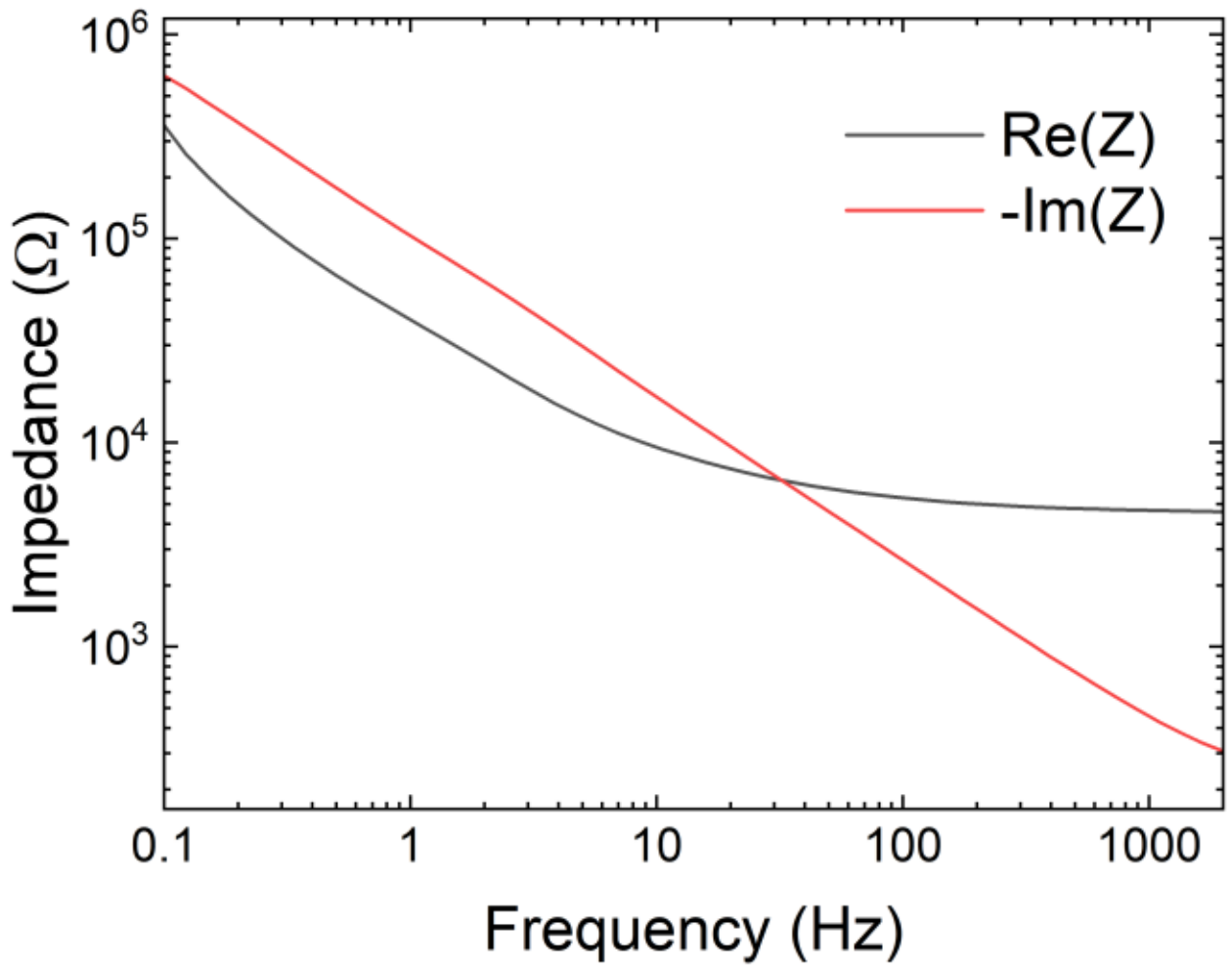

**Fig. S5. Impedance measurement on the Pt surface.** This shows the real and imaginary part of the impedance as a function of the frequency for a Pt surface in DI water on a dedicated test device with a 50 µm gap.

## 3. Supplementary FCS theory: discrete tilted generator and first-passage cargo ensemble

### 3.1 Discrete tilted generator

Full Counting Statistics (FCS) is commonly used in mesoscopic transport to extract the different cumulants describing the statistics of the charge carriers[26,27]. Here we want to show that it can be applied to an electrochemical system, where Ferrocene (Fc) molecule act as room temperature single electron shuttles (Fig. 2A).

To apply FCS to the electrochemical reaction of Ferrocene in a microgap, we consider a discrete model where the space between the two electrodes (source and drain) is divided in N sites (Fig. 2B). In each site, the Fc molecule can be in oxidized or reduced state but can switch between these states only at the left or right electrodes by receiving or giving away an electron. This effectively makes the system into a 2N states loop.

The state of the system at a time $t$ can be represented by a 2N vector $p(t)$ which gives the probability of being in each site. Since Fc molecules do not interact with each other, we consider that there is only one particle in the system and the evolution equation can be written:

$$\delta_t p(t,\chi) = W(\chi)p(t,\chi) \tag{S1}$$

Where $W$ is the rates matrix (Fig 2C) describing the system's evolution and $\chi$ is the counting field introduced to obtain the current statistics. It is used to count electrons exchanged at the left electrode between site 1 (Fc at the left electrode) and site 2 (Fc+ at the left electrode).

The FCS theory gives the link between the cumulants of different orders and $\mu(\chi)$, the dominant eigenvalue of the rates matrix $W$ through the general equation for the cumulants:

$$\kappa_n = \left.\frac{\delta^n \mu(\chi)}{\delta(i\chi)^n}\right|_{\chi=0} \tag{S2}$$

From these equations we can derive the analytical expressions for the current (1$^{st}$ cumulant) and the noise (2$^{nd}$ cumulant) at different values of $N$. When we calculate $\kappa_n$ for a given $N = i$, we need to solve a $2N$ equation system. Once this is done, we have an expression for the moments in the form:

$$\kappa_n(N=i) = \frac{\sum_k P_k(i)\Gamma_D^{a_k}\Gamma_L^{b_k}\Gamma_{LB}^{c_k}\Gamma_R^{d_k}\Gamma_{RB}^{e_k}}{\sum_j P_j(i)\Gamma_D^{a_j}\Gamma_L^{b_j}\Gamma_{LB}^{c_j}\Gamma_R^{d_j}\Gamma_{RB}^{e_j}} \tag{S3}$$

In other words, the moments are the ratio between two sums of polynomials in *N* which are multiplied by a product of powers of the rates. Here $P_k(i)$ is a polynomial in *N* evaluated in $N = i$ and the $a_k$, $b_k$, $c_k$, $d_k$ and $e_k$ are the power to which the rates are elevated. Note that $a_k + b_k + c_k + d_k + e_k = c^{te}$. Same goes for the denominator but the constant is different.

If we calculate the moments for different values of $N$, then we will have the values of each polynomial for these values of *N*. We can then use these values to fit the polynomials to find the general expression. If $d$ is the degree of the biggest polynomials, then we need to evaluate the polynomials for at least *d+1* values of *N* to make the fit. Once the fit is done for every polynomial, we can inject the general expression of the polynomial in the expression of the moment to get the general expression for any *N*.

We then take the limit of this general expression when N goes to infinity to reach the continuous limit. Finally, after being rescaled to take into account the discretization of space and time, the rates are replaced by the Butler-Volmer rates for the electronic transfer rates and the Fick rate for the diffusion rates to obtain the analytical expressions for the current and noise as a function of the applied voltages.

We use the *sympy* Python library to solve symbolic equations. This library, similar to the Mathematica software has many build-in solving or simplification functions. These build-in functions work fine to solve the 2N×2N matrix system for the first moment, but they struggle to solve the system for the second moment for large values of N. We therefore had to write custom solving functions to be able to solve the system for larger N.

The results of the FCS calculations using this custom symbolic calculation code give the following formula for first and second cumulants in the continuous limit ($N \to \infty$) after taking into account the rescaling of the rates:

$$\kappa_1 = \frac{\lambda(\Gamma_L\Gamma_R - \Gamma_{LB}\Gamma_{RB})}{\Gamma_L\Gamma_R + \Gamma_L\Gamma_{RB} + \Gamma_{LB}\Gamma_R + \Gamma_{LB}\Gamma_{RB} + \lambda(\Gamma_L + \Gamma_{LB} + \Gamma_R + \Gamma_{RB})} \tag{S4}$$

$$\begin{aligned}\kappa_2 = (&3\Gamma_L^2\Gamma_{LB}\Gamma_{RB}^3 + 6\Gamma_L^2\Gamma_{LB}\Gamma_{RB}^2\Gamma_D + 3\Gamma_L^2\Gamma_{LB}\Gamma_{RB}\Gamma_D^2 + 4\Gamma_L\Gamma_{LB}^2\Gamma_{RB}^3 + 8\Gamma_L\Gamma_{LB}^2\Gamma_{RB}^2\Gamma_D + 6\Gamma_L\Gamma_{LB}^2\Gamma_{RB}\Gamma_D^2 + 6\Gamma_L\Gamma_{LB}\Gamma_{RB}^3\Gamma_D + 6\Gamma_L\Gamma_{LB}\Gamma_{RB}^2\Gamma_D^2 \\ &+ \Gamma_{LB}^3\Gamma_{RB}^3 + 2\Gamma_{LB}^3\Gamma_{RB}^2\Gamma_D + 3\Gamma_{LB}^3\Gamma_{RB}\Gamma_D^2 + 2\Gamma_{LB}^2\Gamma_{RB}^3\Gamma_D + 3\Gamma_{LB}\Gamma_{RB}^3\Gamma_D^2) \\ \times &\frac{\Gamma_D}{3(\Gamma_L\Gamma_{RB} + \Gamma_L\Gamma_D + \Gamma_{LB}\Gamma_{RB} + \Gamma_{LB}\Gamma_D + \Gamma_{RB}\Gamma_D)^3}\end{aligned}$$

The current and noise can be obtained by replacing the electron transfer rates by the Butler-Volmer rates and the diffusion rate by the Fick rate. The formulas for the current and the noise are given for simplicity in the case where we consider the right electrode (the cathode) to be grounded, so its potential is much smaller than the standard potential of the Fc/Fc$^+$ couple (Fig. 2D-E).

$$I = ne\kappa_1 = \frac{ADFc_B}{z_{gap}} \frac{1}{1 + e^{-f\eta_L} + \frac{D}{k_0 z_{gap}} e^{-(1-\alpha)f\eta_L}} \tag{S5}$$

$$\begin{aligned}S^I = 2ne^2\kappa_2 &= \frac{2De^2nk_0}{3z} \frac{3D^2e^{f\eta_L} + 2Dk_0 z_{gap} e^{\alpha f\eta_L}(e^{f\eta_L} + 3) + k_0^2 z_{gap}^2 (e^{2f\eta_L} + 4e^{f\eta_L} + 3)}{\left(D + 2k_0 z_{gap} \cosh(\alpha f\eta_L)\right)^3} e^{-\alpha f\eta_L} \\ &= \frac{2e}{3} \frac{ADFc_B}{z_{gap}} f\left(D, k_0 z_{gap}, \eta_L\right)\end{aligned}$$

Where $A$ is the area of the electrodes, $D$ the diffusion constant, $F$ the Faraday constant, $c_B$ the Fc concentration in the microgap, $z_{gap}$ the size of the microgap, $k_0$ the heterogeneous constant, $\alpha$ the electron transfer coefficient and $\eta_L = E_L - E_0$ the overpotential of the anode.

### 3.2 First-passage cargo ensemble

We consider here a continuous-space description of the model described above, which we can solve analytically. This continuous description can be obtained by taking the limit $N \to \infty$ on the number of sites of the discrete model presented above, under a suitable scaling of the microscopic rates. More precisely, set $\varepsilon = \frac{z_{gap}}{N}$ the width of an individual microscopic site, and suppose that as $\varepsilon \to 0$,

$$\Gamma_D \sim \frac{D}{\varepsilon^2}\,, \tag{S6}$$

$$(\Gamma_{ox}^L, \Gamma_{red}^L, \Gamma_{ox}^R, \Gamma_{red}^R) \sim \frac{1}{\varepsilon}(k_{ox}^L, k_{red}^L, k_{ox}^R, k_{red}^R)\,. \tag{S7}$$

The spatial scaling (S6) produces a reflected Brownian motion of diffusion coefficient $D$ in the limit description. It is the usual scaling that gives a limit Brownian motion from a discrete random walk in a lattice (Donsker's theorem). It is therefore compatible with the diffusive motion of the Ferrocene particle. The scaling (S7) of the interfacial electron transfer rates provides an effective boundary description of the Butler-Volmer transfer rates with fixed, finite spatially integrated transfer coefficients $k_{ox}^L, k_{red}^L, k_{ox}^R, k_{red}^R$. Indeed, changing the thickness $\varepsilon$ of the transfer layer preserves the value of the integrated transfer rate. We make this more explicit below.

Let $(Y(t), Z(t))$ be the state of a unique Fc cargo at time $t \geq 0$, where $Y(t) \in \{ox, red\}$ is its redox state, and $Z(t) \in [0, z_{\text{gap}}]$ is its position along the gap, defined as the distance with respect to the left electrode (anode). The spatial coordinate $Z$ evolves as a reflected Brownian motion of diffusion coefficient $D$:

$$dZ(t) = \sqrt{2D}\, dW(t) + d\ell_L(t) - d\ell_R(t), \tag{S8}$$

where $W(t)$ is a standard Brownian motion, and

$$\ell_L(t) = \lim_{\varepsilon\downarrow 0} \frac{D}{\varepsilon} \int_0^t 1_{[0,\varepsilon]}\big(Z(s)\big)\, ds, \qquad \ell_R(t) = \lim_{\varepsilon\downarrow 0} \frac{D}{\varepsilon} \int_0^t 1_{[z_{gap}-\varepsilon,\ z_{gap}]}\big(Z(s)\big)\, ds \tag{S9}$$

are the normalized local time that $Z$ spends up to time $t$ around the left and the right electrode, respectively. As given by (S9), they are the limit of the suitably rescaled time spent at the boundary sites, under the scaling (S6).

The state coordinate $Y$ jumps between the two states $ox, red$ as follows. We write $\bar{y}$ for the opposite state to $y$. Call $N_y^b(t)$ the total number of electron transfer events cumulated by time $t$ through the boundary $b \in \{L, R\}$ (for left/right electrode) that have changed the redox state of the Fc molecule into $y \in \{ox, red\}$. Such transfer events occur at position-dependent transfer rates denoted $\Gamma_y^b(z)$. Mathematically, we can write the counting processes $N_y^b(t)$ as time-changed Poisson processes of the form

$$N_y^b(t) = P_y^b\left(\int_0^t \Gamma_y^b(Z(s))\mathbf{1}_{\{Y(s^-)=\bar{y}\}}\, ds\right), \tag{S10}$$

where the four $P_y^b(t)$ are independent unit-rate Poisson processes, independent of $W(t)$. In particular, we suppose that the transfer rates decay exponentially with the distance of the molecule from the electrode:

$$\Gamma_y^b(z) = \beta\, k_y^b \exp(-\beta d_b(z)) \ , \qquad d_L(z) = z, \qquad d_R(z) = z_{gap} - z, \tag{S11}$$

where $\beta > 0$ is the tunnelling decay ratio, $d_b(z)$ is the distance from electrode $b$ at position $z$, and $k_y^b$ are the effective transfer coefficients introduced in (S7), which have units of length/time. Notice that if $\beta$ is sufficiently large with respect to $z_{gap}$, the integrated transfer rate across the nanogap gives indeed fixed effective transfer rates $k_y^b$:

$$\int_0^{z_{gap}} \Gamma_y^b(z)\ dz = k_y^b\left(1 - e^{-\beta\, z_{gap}}\right) \xrightarrow{\beta z_{gap} \gg 1} k_y^b\,. \tag{S12}$$

In general, any transfer rate of the form $\Gamma_y^b(z) = \beta\, k_y^b\, r(\beta\, d_b(z))$, where $r$ is a non-negative function such that $\int_0^\infty r(z)dz = 1$, would yield the same asymptotic. In particular, setting $r_y^b(u) = \mathbf{1}_{\{u<1\}}$, i.e., forcing the transfer to occur at fixed rate only in a layer of thickness $\varepsilon = \beta^{-1}$ from the respective electrode, as in (S7), would also lead to the same finite interfacial kinetics. Hereafter, $r_y^b(u) = e^{-u}$.

Using (S11) and since $e^{-v} = \int_v^\infty e^{-u}du$, notice that

$$\int_0^t \Gamma_y^b(Z(s))\, ds = \beta\, k_y^b \int_0^\infty e^{-u}\int_0^t \mathbf{1}_{\left\{d_b(Z(s))<\frac{u}{\beta}\right\}}\, ds\, du \longrightarrow \frac{k_y^b}{D}\ell_b(t), \tag{S13}$$

as $\beta \to \infty$ (dominated convergence). Thus, as $\beta \to \infty$, the limit transfer counting processes defined in (S10) are of the form

$$N_y^b(t) = P_y^b\left(\frac{k_y^b}{D}\int_0^t \mathbf{1}_{\{Y(s^-)=\bar{y}\}}\, d\ell_b(s)\right). \tag{S14}$$

Transfer events occur at finite rates per unit boundary local time. Starting at boundary $b$ in state $y$, the probability of reaching the opposite electrode before a charge transfer is $\left(1 + 1/\Lambda_{\bar{y}}^b\right)^{-1} > 0$ (see for example, (*49*), where the constant $\Lambda_y^b = D/(z_{gap}k_y^b)$ is the dimensionless rate constant. In particular, we are not imposing a Nernstian regime.

We are interested in the statistics of the net counting of electrons transferred across the anode,

$$\mathcal{N}(t) = N_{ox}^L(t) - N_{red}^L(t), \tag{S15}$$

fully characterized by its characteristic function,

$$S_t(\chi \mid y, z) = \mathbb{E}\left[e^{i\chi\,\mathcal{N}(t)} \mid Y(0) = y,\ Z(0) = z\right], \qquad \chi \in \mathbb{R}. \tag{S16}$$

It is well known that Feynman-Kac formula identifies $S_t(\chi \mid y, z)$ as solution of the Partial Differential Equation in $H^2(0, z_{gap})$

$$\partial_t S_t(\chi \mid y, z) = \mathcal{L}_\chi[S_t(\chi \mid \cdot)](y, z), \qquad \partial_z f(y, 0) = \partial_z f(y, z_{gap}) = 0, \qquad S_0(\chi \mid y, z) = 1, \tag{S17}$$

where the Neumann boundary condition imposes reflective boundaries at each electrode and, setting $\bar{y} = ox$ if $y = red$, and vice versa $\bar{y} = red$ if $y = ox$,

$$\mathcal{L}_\chi[f](y, z) = D\partial_{zz} f(y, z) + \Gamma_{\bar{y}}^R(z)\big(f(\bar{y}, z) - f(y, z)\big) + \Gamma_{\bar{y}}^L(z)\left(e^{\left(1_{\{y=red\}} - 1_{\{y=ox\}}\right)\mathrm{i}\,\chi} f(\bar{y}, z) - f(y, z)\right) \tag{S18}$$

is the tilted generator, analogous to the tilted matrix $W(\chi)$ in the discrete system (recall (S1) and Fig. 2C), but now in infinite dimension. The exponential tilts have again the same role. The discrete hopping between sites is now replaced by the Laplacian term $D\partial_{zz} f(y, z)$ that generates the Brownian motion (S8). Under standard spectral assumptions, for $\chi$ sufficiently close to 0, there are Perron eigenfunctions $p_\chi$, $q_\chi$ associated to a principal eigenvalue $\mu(\chi)$ such that $\mu(0) = 0$ and

$$\mathcal{L}_\chi[p_\chi] = \mu(\chi)p_\chi, \qquad \mathcal{L}_\chi^*[q_\chi] = \mu(\chi)q_\chi, \qquad \int_0^{z_{gap}} p_\chi\,.\,q_\chi\, dz = 1, \qquad \int_0^{z_{gap}} q_\chi.1\, dz = 1\,, \tag{S19}$$

where $\mathcal{L}_\chi^*$ is the formal transpose of $\mathcal{L}_\chi$, so that as $t \to \infty$, the characteristic function admits the asymptotic expansion

$$S_t(\chi \mid y, z) = e^{\mu(\chi)t}\, p_\chi(y, z) + O\left(e^{(\mu(\chi)-\eta_\chi)t}\right), \qquad \text{for some } \eta_\chi > 0. \tag{S20}$$

In particular, independently of the initial values $Y(0) = y,\ Z(0) = z$, we have

$$\lim_{t\to\infty}\frac{1}{t}\log S_t(\chi\mid y,z) = \mu(\chi), \tag{S21}$$

where $\mu(\chi)$ is called the (second) scaled cumulant generating function (SCGF). Then, the long-time asymptotic cumulants can be obtained through successive differentiation of $\mu(\chi)$ at $\chi = 0$. That is, if $\kappa_n(t)$ is the $n$-th cumulant of $\mathcal{N}(t)$, we have

$$\lim_{t\to\infty}\frac{1}{t}\kappa_n(t) = (-i)^n \frac{d^n}{d\chi^n}\mu(\chi)\Big|_{\chi=0}. \tag{S22}$$

The eigenvalue problem (S19) is not easy to solve in full generality. The discrete-state process considered in 3.1. allowed to reduce it to a linear algebra problem. Here, the limit $\beta\to\infty$ allows to reduce (S19) to the simpler ODE

$$\begin{cases} D\,\boldsymbol{p}_\chi''(z) = \mu(\chi)\boldsymbol{p}_\chi(z), \qquad 0<z<z_{gap}, \\ D\boldsymbol{p}_\chi'(0) + B_L(\chi)\boldsymbol{p}_\chi(0) = 0 \\ -D\boldsymbol{p}_\chi'(z_{gap}) + B_R\boldsymbol{p}_\chi(z_{gap}) = 0 \end{cases} \tag{S23}$$

where $\boldsymbol{p}_\chi(z) = \left(p_\chi(ox,z), p_\chi(red,z)\right)^{\top}$, and

$$B_L(\chi) = \begin{pmatrix} -k_{red}^L & k_{red}^L e^{-i\chi} \\ k_{ox}^L e^{i\chi} & -k_{ox}^L \end{pmatrix}, \qquad B_R = \begin{pmatrix} -k_{red}^R & k_{red}^R \\ k_{ox}^R & -k_{ox}^R \end{pmatrix}. \tag{S24}$$

***Proof of (S23):*** Using the limit form (S14) and applying Ito's formula with jumps (see (*50*) for example) to

$$F(t) = e^{i\chi N(t)-\mu(\chi)\,t}\,p_\chi(Y(t),Z(t)) \tag{S25}$$

we get

$$dF(t) = d\mathcal{M}(t) + e^{i\chi\mathcal{N}(t^-)-\mu(\chi)t}\,\Big(\left[D\boldsymbol{p}_\chi''(Z(t)) - \mu\boldsymbol{p}_\chi(Z(t))\right]^{Y(t^-)} dt$$
$$+\left[\boldsymbol{p}_\chi'(0) + \frac{1}{D}B_l(\chi)\boldsymbol{p}_\chi(0)\right]^{Y(t^-)} d\ell_L(t) + \left[-\boldsymbol{p}_\chi'(z_{gap}) + \frac{1}{D}B_r\boldsymbol{p}_\chi(z_{gap})\right]^{Y(t^-)} d\ell_R(t)\Big), \tag{S26}$$

where $[\boldsymbol{p}(z)]^y = p(y,z)$, and $\mathcal{M}(t)$ collects the Brownian and compensated-jump martingales. Thus, the drift and the local-time terms of $F(t)$ vanish precisely when $p_\chi$ satisfies (S23), so that we have the Feynman-Kac representation

$$\mathbb{E}\left[e^{i\chi N(t)}p_\chi(Y(t),Z(t)) \mid Y(0)=y, Z(0)=z\right] = e^{\mu(\chi)\,t}p_\chi(\mathrm{y},\mathrm{z}). \tag{S27}$$

Moreover, as stated in (S1),

$$\mu(\chi) = \lim_{t\to\infty}\frac{1}{t}\log\mathbb{E}\left[e^{i\chi N(t)} \mid Y(0)=y, Z(0)=z\right], \qquad \mu(0)=0.$$

We can then correctly identify the solution of (S23) to the right eigenvector of (S19).

■

Now, set $k_{ox}^R = 0, k_{red}^R\to\infty$. The right boundary condition in (S23) becomes

$$p_\chi(ox,z_{gap}) = p_\chi(red,z_{gap}) \quad\text{and}\quad \frac{d}{dz}p_\chi(red,z)\Big|_{z=z_{gap}} = 0. \tag{S28}$$

Set $\phi = \log(\Lambda_{red}^L/\Lambda_{ox}^L)$ and $\Lambda = \sqrt{\Lambda_{ox}^L\Lambda_{red}^L}$, so that, the effective transfer rates at the left electrode can be written in the symmetric Butler-Volmer form

$$k_{ox}^L = k_0\,e^{\phi/2}, \qquad k_{red}^L = k_0\,e^{-\phi/2}, \tag{S29}$$

where $k_0 = (D/z_{gap})/\Lambda$ is the standard heterogeneous electron transfer rate constant, with $\Lambda$ the dimensionless rate constant introduced in Eq. (1), and $\phi = f(V-E_0)$ in the notations of Eq. (2). Set also

$$\rho_{ox}^L = \frac{1}{1+e^{-\phi}+\Lambda e^{-\frac{\phi}{2}}}, \qquad \Delta = \sqrt{\Lambda^2{\rho_{ox}^L}^2 + 4\rho_{ox}^L(1-\rho_{ox}^L)}, \qquad \vartheta = \frac{\Lambda}{2}(\Delta - \Lambda\rho_{ox}^L). \tag{S30}$$

to simplify the following expressions. In particular, $\rho_{ox}^L$ is the oxidized fraction at the anode as described by Eq. (1) (we show this explicitly below). Now we show that, for $\chi$ around 0, the SCGF $\mu(\chi)$ is determined by the fixed point of

$$\mu(\chi) = \frac{D}{z_{gap}^2}\left[\operatorname{arcsinh}\sqrt{\frac{\rho_{ox}^L(e^{i\chi}-1)}{1+\vartheta(u(\chi)\coth u(\chi) - 1)}}\right]^2, \qquad u^2(\chi) = \frac{z_{gap}^2}{D}\mu(\chi), \qquad \mu(0)=0, \tag{S31}$$

with $u\coth u$ extended continuously by 1 at $u=0$.

***Proof of (S31):*** Solving the second order ODE (S23) with the right boundary conditions replaced by (S28) gives

$$p_\chi(red,z) = A\cosh\left(u(\chi)\left(1-\frac{z}{z_{gap}}\right)\right), \qquad p_\chi(ox,z) = p_\chi(red,z) + B\,\frac{\sinh\left(u(\chi)\left(1-\frac{z}{z_{gap}}\right)\right)}{u(\chi)} \tag{S32}$$

Substituting into the left boundary condition in (S23) clears out the constants $A$, $B$ and yields

$$\cosh^2 u(\chi) + e^{-\phi}\sinh^2 u(\chi) + \Lambda e^{-\phi/2} u(\chi)\sinh u(\chi)\cosh u(\chi) = e^{i\chi} \tag{S33}$$

Differentiating implicitly at $\chi = 0$ gives $\mu'(0) = i\,D\,\rho_{ox}^L/z_{gap}^2$. In particular, we have indeed $\rho_{ox}^L = -i\left(z_{gap}^2/D\right)\mu'(0) = I/I_{lim}$. Then, since $\rho_{ox}^L\,\Lambda e^{-\phi/2} = \vartheta$, Eq. (S33) can be rewritten as

$$\left(1+\vartheta\left(u(\chi)\coth u(\chi) - 1\right)\right)\sinh^2[u(\chi)] = \rho_{ox}^L(e^{i\chi}-1). \tag{S34}$$

which proves (S31).

■

Finally, coming back to (S21), the cumulant generating function of $\mathcal{N}(t)$ scales as $\mu(\chi)t + O(1)$ as $t\to\infty$. Writing $\langle n\rangle_t = -i\mu'(0)t$ for the mean electron count in a time window of size $t$, we have then, as $t\to\infty$,

$$S_t(\chi) = \frac{\langle n\rangle_t}{\rho_{ox}^L}\left(\operatorname{arcsinh}\sqrt{\frac{\rho_{ox}^L(e^{i\chi}-1)}{1+\frac{\Lambda}{2}(\Delta-\Lambda\rho_{ox}^L)(u(\chi)\coth u(\chi)-1)}}\right)^2 + O(1), \tag{S35}$$

independently of the initial conditions.
In the Nernstian regime where $\Lambda \ll 1$, (S35) directly becomes Eq. (5) of the main text:

$$S(\chi) = \frac{\langle n\rangle_t}{\rho_{ox}^L}\left(\operatorname{arcsinh}\sqrt{\rho_{ox}^L(e^{i\chi}-1)}\right)^2 \tag{5}$$

We give an alternative derivation based on a renewal decomposition, which is useful to decouple the contributions of Brownian diffusion and redox cycling to the form of the characteristic function, and in particular, to the noise reduction in absence of multi-particle exclusion. By (S14), let us represent the redox dynamics at electrode $b$ by Poisson resampling events of intensity $(k_b^{ox}+k_b^{red})/D$ per unit boundary local time. At each event at boundary $b$, independently from the time elapsed, the molecule is oxidized with probability

$$\wp_b = \frac{k_b^{ox}}{k_b^{ox}+k_b^{red}}, \tag{S36}$$

and otherwise reduced with probability $1-\wp_b$. Let us initialize the Fc particle at the left boundary ($b = L,\ z = 0$), at time $\tau_0^L = 0$, immediately after a redox event, and for all $n = 0,1,2\ ...$, define $\tau_n^R$ as the time of the first resampling at the right boundary after $\tau_n^L$, and $\tau_{n+1}^L$ as the time of the first left resampling after $\tau_n^R$. The duration of a complete cycle between two anode resamplings is then

$$T_n = \tau_{n+1}^L - \tau_n^L. \tag{S37}$$

Let $Y_n^L = Y(\tau_n^{R\,-})$ the redox state assigned at the last left resampling before the right resampling time $\tau_n^L$, and $Y_n^R = Y(\tau_{n+1}^{L\,-})$ the state assigned at the last left resampling before the right resampling time $\tau_n^R$. The pairs $(Y_n^L, Y_n^R)$ are independent across cycles and independent of the cycle duration, and by (S36), for all $n$,

$$Y_n^b = \begin{cases} ox, & \text{with probability } \wp_b \\ red, & \text{with probability } 1-\wp_b \end{cases} \tag{S38}$$

Notice that $Y_n^R$ is the incoming redox state at the left electrode at the resampling time $\tau_{n+1}^L$. Define the $n$-th contribution to the total count by $Q_n = 1_{Y_n^L=ox} - 1_{Y_n^R=ox}$. The net counting in (S15) is then equivalent to the renewal counting process

$$\mathcal{N}(\tau_n^L) = \sum_{m=0}^{n-1} Q_m + \left(1_{Y(\tau_n^L)=ox} - 1_{Y(0)=ox}\right), \tag{S39}$$

where the disagreement at the boundary is negligible in the long-time limit. Both counts have the same limit SCGF. Let

$$F(t) = \mathbb{P}(T_n \le t), \qquad g(s) = \int_0^\infty e^{-st}\,F(dt), \qquad G(\chi) = \mathbb{E}[e^{\chi Q_n}]. \tag{S40}$$

Conditioning on the first cycle and using independence of its duration and count contributions gives, for any $\mu,\chi\in\mathbb{R}$,

$$e^{-\mu t}S_t(\chi) = e^{-\mu t}\left(1-F(t)\right) + e^{-\mu t}G(\chi)\int_0^t S_{t-s}(\chi)\,F(ds). \tag{S41}$$

Choosing $\mu$ so that $G(\chi)e^{-\mu s}F(ds)$ is a probability kernel, the Key Renewal Theorem (see for example (*51*)) gives $e^{-\mu t}S_t \to C_\chi > 0$ and hence the desired limit (S21). This condition is determined by the scalar equation

$$G(\chi)g\big(\mu(\chi)\big) = 1, \qquad \mu(0) = 0. \tag{S42}$$

It remains to calculate the two ingredients: the redox count moment generating function $G$ and the cycle-duration transform $g$. From (S38), the moment-generating function of $Q_n$ is

$$G(\chi) = \mathbb{E}[e^{\chi\, Q_n}] = (1 - \wp_L + \wp_L\, e^{\chi})(1 - \wp_R + \wp_R\, e^{-\chi}). \tag{S43}$$

It remains to determine the cycle duration law. For a particle started at the opposite electrode, let $\zeta_b$ the first resampling time at electrode $b$. Equivalently, $\zeta_b$ is the time when the boundary local time first exceeds an independent exponential threshold of rate $k_b^{ox} + k_b^{red}/D$. A classical scalar Brownian first-passage time calculation gives (see for example (*43*)):

$$\phi_b(s) = \mathbb{E}[e^{-s\,\zeta_b} | Z_0 = opposite\ boundary] = \left(\cosh u + \frac{D}{z_{gap}(k_b^{ox} + k_b^{red})} u(s) \sinh u(s)\right)^{-1}, \qquad u^2(s) = \frac{z_{gap}^2 s}{D} \tag{S44}$$

By the strong Markov property each one-way travels is independent of the others and therefore

$$g(s) = \mathbb{E}[e^{-s\, T_n}] = \operatorname{sech}^2 u(s) \prod_{b\in\{L,R\}} \left(1 + \frac{D}{z_{gap}(k_b^{ox} + k_b^{red})} u(s) \tanh u(s)\right)^{-1} \tag{S45}$$

Substituting (S43) and (S45) into the renewal condition (S42) gives

$$(1 - \wp_L + \wp_L e^{\chi})(1 - \wp_R + \wp_R e^{-\chi}) = \prod_{b\in\{L,R\}} \left(\cosh u(\chi) + \frac{D\, u(\chi) \sinh u(\chi)}{z_{\text{gap}}(k_b^{ox} + k_b^{red})}\right), \qquad u^2(\chi) = \frac{z_{\text{gap}}^2 \mu(\chi)}{D}, \qquad \mu(0) = 0. \tag{S46}$$

This is equivalent to the scalar equation obtained from the Robin eigenproblem (S23). This formulation however separates the contributions of each ingredient to the full distribution: the term $\operatorname{sech}^2 u$ comes from Brownian round-trip first-passage times, while the factors involving $k_b^{ox} + k_b^{red}$ describe the additional time required by redox resampling events. In particular, as $k_b^{ox}, k_b^{red} \to \infty$, with $\wp_b$ fixed, the additional terms vanish and one obtains the equation

$$G(\chi) \operatorname{sech}^2\left(z_{gap}\sqrt{\mu(\chi)/D}\right) = 1, \tag{S47}$$

and therefore, the more familiar form

$$\mu(\chi) = \frac{D}{z_{gap}^2}\left(\operatorname{arcsinh}\sqrt{G(\chi) - 1}\right)^2. \tag{S48}$$

We can finally substitute $\chi$ by $i\,\chi$ locally around 0 to come back to the characteristic function formulation.

## 4. Thermal and Poisson reference scales for redox-cycling noise

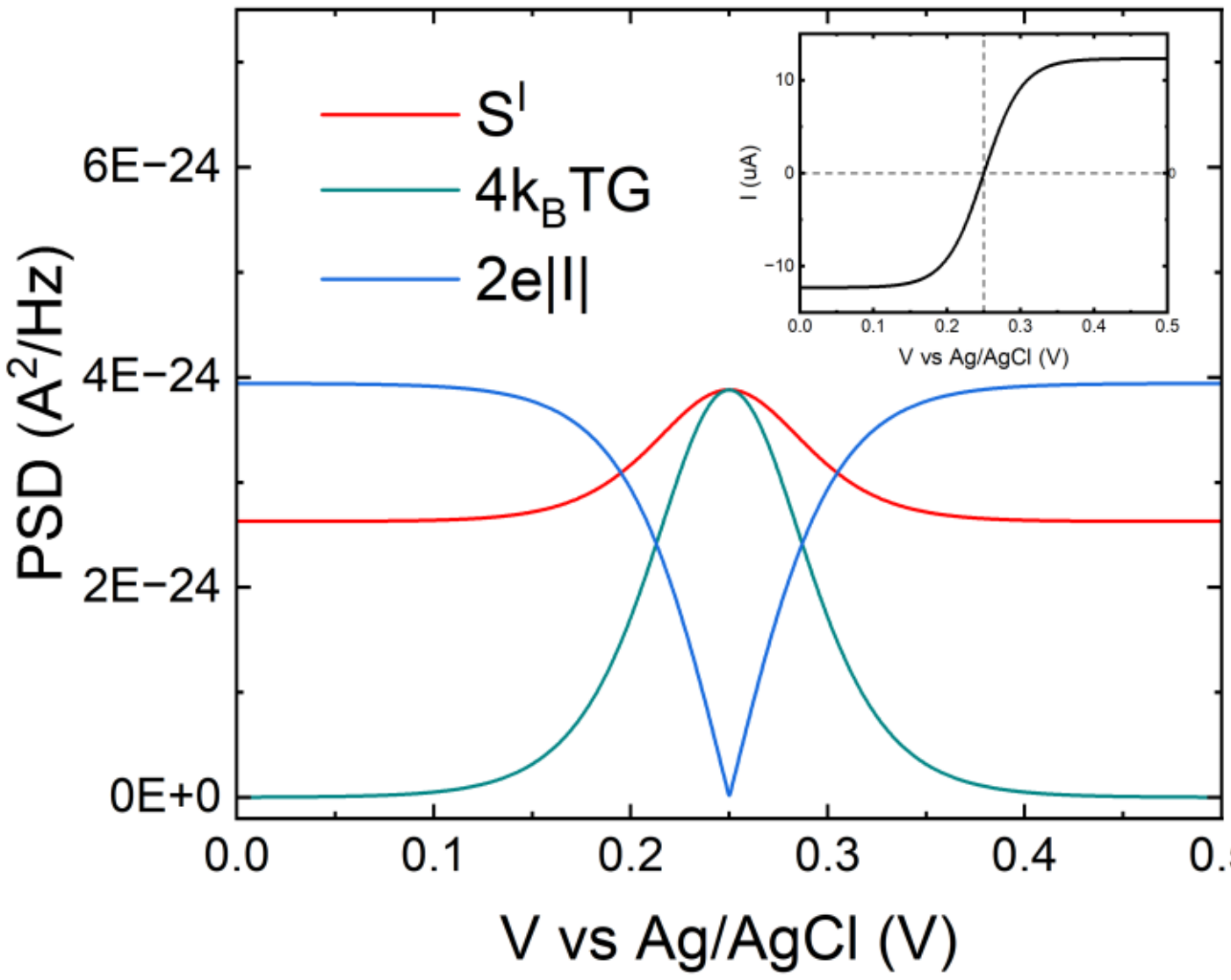

**Fig. S6. Thermal and Poisson reference scales for redox-cycling noise.** Analytical noise $S^I$ in the same condition as Figure 2 of the main text except here the cathode potential $V_R$ is fixed at the formal potential $E_0$ instead of being much smaller. The current noise from Equation S5 is compared with $4k_BTG$ and $2e|I|$, as a function of the swept anodic potential $V$ versus Ag/AgCl. The net current vanishes at $V = V_R = E_0$, while forward and reverse electron-transfer events still produce the equilibrium noise equal to $4k_BTG$. Inset: current, which is antisymmetric about $E_0$ and bounded by the diffusion-limited plateau.

## 5. $P(n)$ distributions at different $\rho_{ox}^{L}$ in the Nernstian and non-Nernstian regime

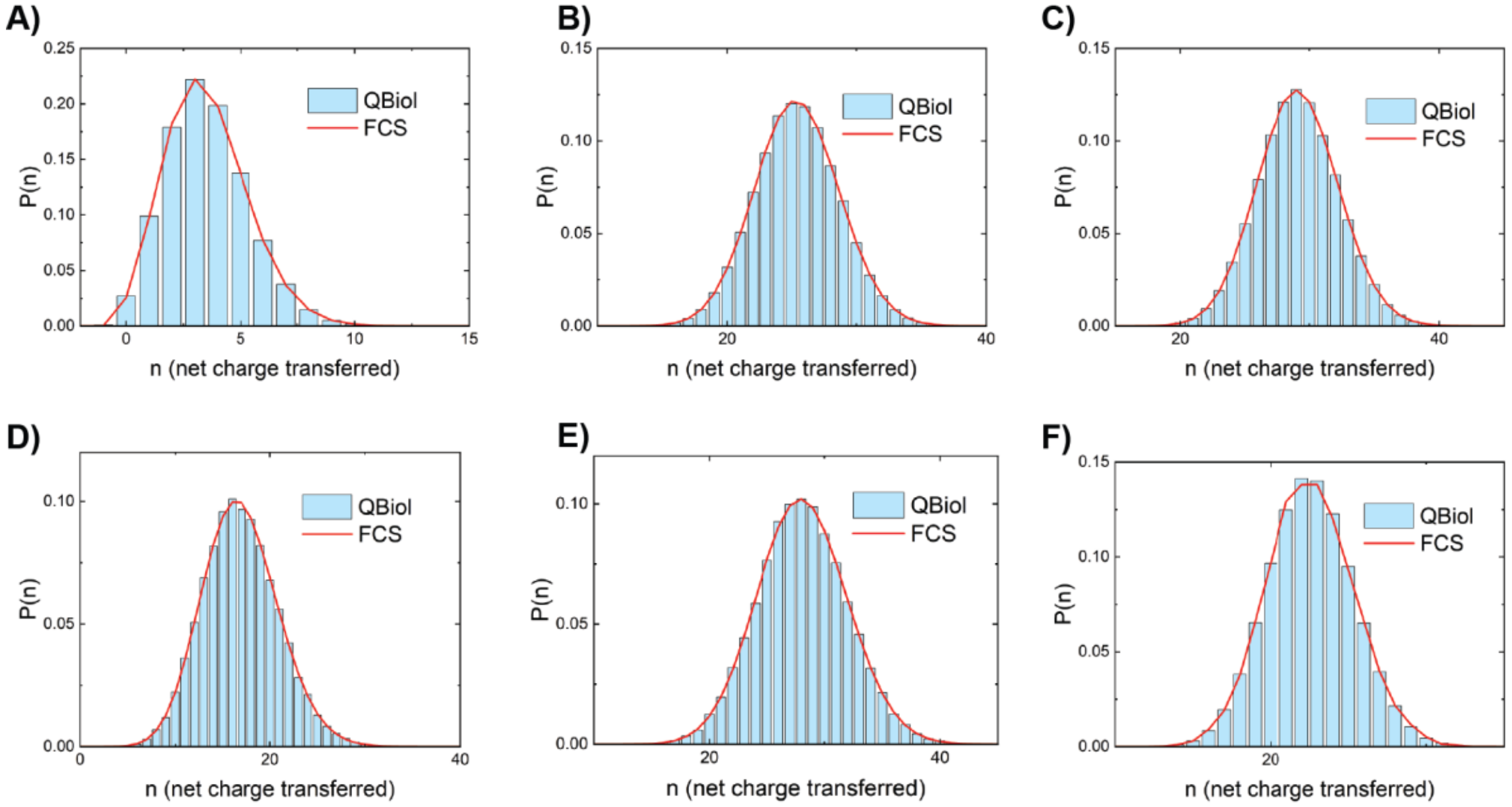

**Fig. S7. Full counting distributions at different in the Nernstian and non-Nernstian regimes at different $V_L$ with $V_R = -0.05$ V.** QBiol histograms are compared with the main text Equation 5 in the Nernstian case, with $z_{gap} = 5$ µm: (**A**) $V_L = 0.2$ V, (**B**) $V_L = 0.3$ V and (**C**) $V_L = 0.4$ V; and with supplementary Equation S35 in the non-Nernstian case with $z_{gap} = 60$ nm: (**D**) $V_L = 0.2$ V, (**E**) $V_L = 0.3$ V and (**F**) $V_L = 0.4$ V.

## 6. Higher order moments

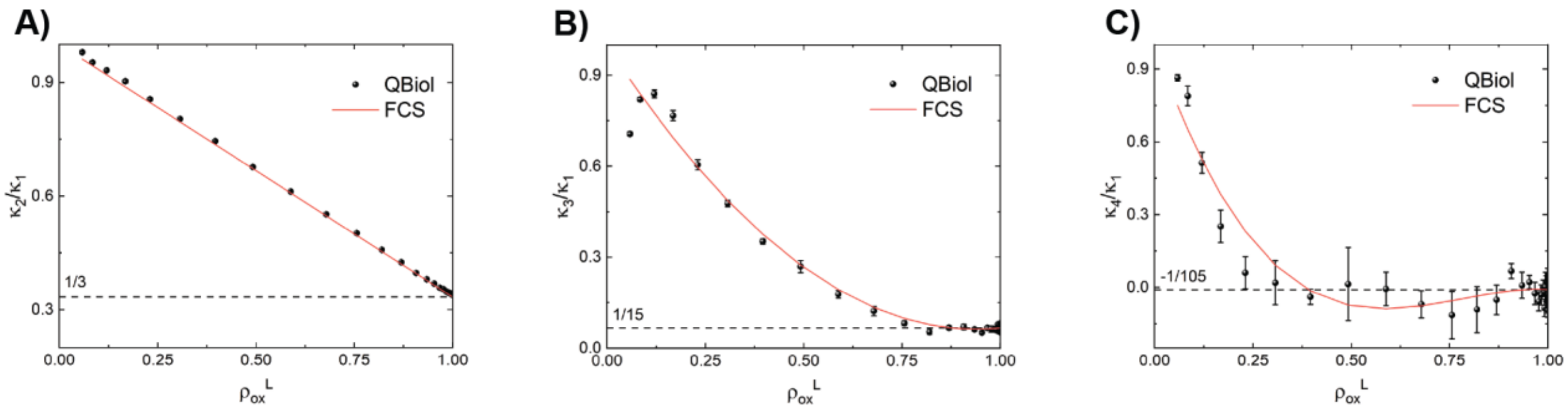

**Fig. S8. Fano factor and higher order cumulants as a function of $\rho_{ox}^{L}$.** The Fano factor and higher cumulants were obtained by the QBiol simulation for a 5 µm gap and compared to the theoretical value obtained by the generating function (Eq. 5). At high bias (i.e. when $\rho_{ox}^{L} \to 1$) they fall from their Poissonian values toward $1/3$, $1/15$ and $-1/105$ respectively, the universal values for diffusive systems. In these 3 cases, an analytical model is obtained. Eq. 4 for the Fano; 3rd cumulant $\kappa_3/\kappa_1 = 1 - 2\rho_{ox}^{L} + \frac{16}{15}{\rho_{ox}^{L}}^{2}$; 4th cumulant: $\kappa_4/\kappa_1 = 1 - \frac{14}{3}\rho_{ox}^{L} + \frac{32}{5}{\rho_{ox}^{L}}^{2} - \frac{96}{35}{\rho_{ox}^{L}}^{3}$